\documentclass[twocolumn,aps,showpacs,superscriptaddress,floatfix]{revtex4-1}

\usepackage{graphicx}
\usepackage{dcolumn}
\usepackage{bm}
\usepackage{color}
\usepackage{amsmath} 
\newcommand{\angstrom}{\text{\normalfont\AA}}
\usepackage{multirow}
\usepackage{epsfig}
\usepackage{dcolumn}

\newcommand{\srni}{SrNi$_2$V$_2$O$_8$}
\def\be{\begin{equation}}
\def\ee{\end{equation}}

\usepackage{color}
\usepackage[colorlinks,breaklinks,bookmarks=true,citecolor=blue,linkcolor=red,urlcolor=blue]{hyperref}

\begin{document}

\title{Consequences of critical interchain couplings and anisotropy on a Haldane-chain}

\author{A. K. Bera}
\email{anup.bera@helmholtz-berlin.de}
\affiliation{Helmholtz-Zentrum Berlin f{\"u}r Materialien und Energie, D-14109 Berlin, Germany}

\author{B. Lake}
\affiliation{Helmholtz-Zentrum Berlin f{\"u}r Materialien und Energie, D-14109 Berlin, Germany}
\affiliation{Institut f{\"u}r Festk{\"o}rperphysik, Technische Universit{\"a}t Berlin, Hardenbergstra{\ss}e 36, D-10623 Berlin, Germany}

\author{A. T. M. N. Islam}
\affiliation{Helmholtz-Zentrum Berlin f{\"u}r Materialien und Energie, D-14109 Berlin, Germany}

\author{O. Janson}
\affiliation{Max-Planck-Institut f{\"u}r Chemische Physik fester Stoffe, N{\"o}thnitzer Stra{\ss}e 40, 01187 Dresden, Germany}%

\author{H.~Rosner}
\affiliation{Max-Planck-Institut f{\"u}r Chemische Physik fester Stoffe, N{\"o}thnitzer Stra{\ss}e 40, 01187 Dresden, Germany}%

\author{A. Schneidewind}
\affiliation{J{\"u}lich Centre for Neutron Science JCNS, Forschungszentrum J{\"u}lich GmbH, Outstation at MLZ, D-85747 Garching, Germany}

\author{J. T. Park}
\affiliation{Heinz Maier-Leibnitz Zentrum, TU M{\"u}nchen, D-85747 Garching, Germany}

\author{E. Wheeler}
\affiliation{Institut Laue-Langevin, Boite Postale 156, 38042 Grenoble Cedex, France}

\author{S. Zander}
\affiliation{Helmholtz-Zentrum Berlin f{\"u}r Materialien und Energie, D-14109 Berlin, Germany}

\date{\today}

\begin{abstract}
Effects of interchain couplings and anisotropy on a Haldane chain have been investigated by single crystal inelastic neutron scattering and density functional theory (DFT) calculations on the model compound \srni. Significant effects on low energy excitation spectra are found where the Haldane gap ($\Delta_0 \approx 0.41J$; where $J$ is the intrachain exchange interaction) is replaced by three energy minima at different antiferromagnetic zone centers due to the complex interchain couplings. Further, the triplet states are split into two branches by single-ion anisotropy. Quantitative information on the intrachain and interchain interactions as well as on the single-ion anisotropy are obtained from the analyses of the neutron scattering spectra by the random phase approximation (RPA) method. The presence of multiple competing interchain interactions is found from the analysis of the experimental spectra and is also confirmed by the DFT calculations. The interchain interactions are two orders of magnitude weaker than the nearest-neighbour intrachain interaction $J$ = 8.7~meV. The DFT calculations reveal that the dominant intrachain nearest-neighbor interaction occurs via nontrivial extended superexchange pathways Ni--O--V--O--Ni involving the empty $d$ orbital of V ions. The present single crystal study also allows us to correctly position \srni\ in the theoretical $D$-$J_{\perp}$ phase diagram [T. Sakai and M. Takahashi, Phys. Rev. B 42, 4537 (1990)] showing where it lies within the spin-liquid phase.
\end{abstract}


\maketitle

\section {Introduction}
\begin{figure}
\includegraphics[trim=0.1cm 0cm 0cm 0cm, clip=true, width=86mm]{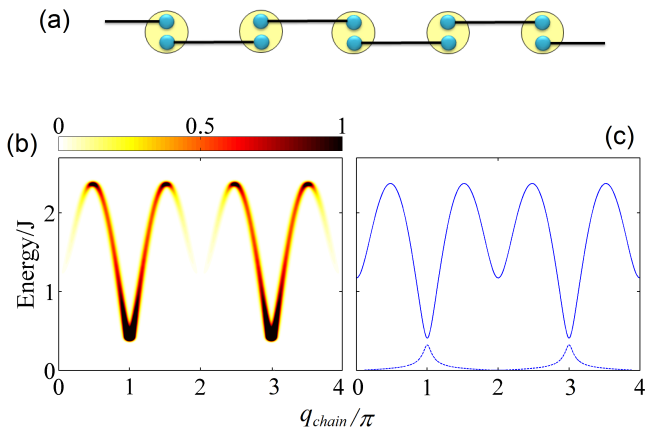} 
\caption{\label{Fig:Haldane}(Color online) (a) A schematic representation of the singlet ground state of a Haldane chain in terms of a valence bond solid state. (b) A schematic representation (color plot) of the one-magnon excitation spectrum of an isolated Haldane chain over two Brillouin zones. (c) The dispersion relation (solid line) and the intensity profile (dashed line at the bottom) of the excitation spectrum in (b).}
\end{figure}

Spin-1 Heisenberg antiferromagnetic (AFM) chains or Haldane chains are of current interest due to their novel magnetic properties \cite{AffleckJPCM.1.3047,SmirnovJETP.105.861}. In such systems, the magnetic ions interact with their nearest neighbors in only one direction (1D) and the long-range magnetic order is suppressed even at $T$ = 0~K by strong quantum spin fluctuations. They have exceptional dynamical properties with exponential decay of spin correlation functions. Haldane predicted a unique many body singlet ground state and gapped magnetic excitations \cite{HaldanePRL.50.1153, HaldanePLA.93.464}. The energy gap between the singlet ground state and the 1$^{st}$ excited state (triplet) is known as the Haldane gap. This is in contrast to a gapless continuum of multi-spinon excitations ($S$ = 1/2) of its half-integer counterpart, the Spin-1/2 Heisenberg uniform AFM chain \cite{HaldanePRL.50.1153, HaldanePLA.93.464}. The singlet ground state of the Haldane chain can be visualized in a simple way similar to the valence bond solid (VBS) state.  Here, the VBS state is constructed by representing each S =1 spin as two separate S =1/2 spins which form a singlet with their counterparts on the two neighboring sites \cite{Affleck.PRL.59.799} as schematically shown in Fig.~\ref{Fig:Haldane}(a). Each exchange bond carries exactly one valence bond, and the periodicity of the underlying lattice remains unbroken.  
 
The excitation spectra of an isolated Haldane chain were investigated in great detail by several theoretical methods including quantum monte-carlo (QMC) \cite{TakahashiPRL.62.2313}, Exact diagonalization (ED) \cite{Golinelli.JPCM.5.1399} and density matrix renormalization group (DMRG) \cite{WhitePRB.77.134437} as well as experimental techniques \cite{Ma.PRL.69.3571,Xu.Science.289.419}. The excitation spectra were well understood as gapped dispersive excitation of magnons ($S$ = 1). The minimum gap of the excitation spectra (Haldane gap) appears at the AFM zone centre ($q_{chain} = \pi$). DMRG calculations estimate the gap value $\Delta_0 = 0.41J$, where $J$ is the intrachain antiferromagnetic coupling \cite{WhitePRB.77.134437}. The one-magnon excitations disperse up to a maximum energy of $\sim$~2.4$J$ at the zone boundary ($q_{chain} = \pi/2$ and 3$\pi/2$). A schematic excitation spectrum of an isolated isotropic Haldance chain is shown in Fig.~\ref{Fig:Haldane}(b). One of the interesting characteristics of the excitation spectrum of Haldane chain is that the dispersion relation is asymmetric about $q_{chain} = \pi/2$ and 3$\pi/2$ [Fig.~\ref{Fig:Haldane}(b) and (c)] since the translational symmetry of the lattice is not broken \cite{Ma.PRL.69.3571,ZaliznyakPRL.87.017202}. This is in contrast to the conventional N{\'e}el antiferromagnets where the unit cell is doubled and $q_{chain} = \pi/2$ and 3$\pi/2$ are of high symmetry points of the spin-wave dispersion relation. The intensity of the excitations [Fig.~\ref{Fig:Haldane}(b) and (c)] is well accounted by the single mode approximation \cite{Ma.PRL.69.3571} which shows a maximum at the zone center and decreases continuously away from $q = \pi$ and disappears rapidly beyond the zone boundary \cite{Ma.PRL.69.3571,Xu.Science.289.419}. In addition to the one-magnon excitations, multi-magnon excitations are also present. Even- and odd-magnon excitations are predicted around the wavevectors $q_{chain} =2n\pi$, and $n\pi$, respectively, where $n$ is an integer \cite{WhitePRB.77.134437}. Thus the dynamics of the isolated Haldane chain are now well understood theoretically and experimentally. Our present research activities aim to explore and understand the influence of interchain interactions and anisotropy on Haldane chains.

The presence of interchain interactions and anisotropy leads to complex behaviors and a rich phase diagram as mapped theoretically by Sakai and Takahashi \cite{SakaiPRB.42.4537}. The Haldane gap can even be suppressed by critical values of interchain interactions and anisotropy which results in three dimensional (3D) long-range magnetic ordering \cite{MorraPRB.38.543}. Experimental challenges are to investigate the effects of interchain interactions and anisotropy on a model compound, to verify the theoretical predictions as well as to explore the critical properties. In this context, among the experimentally reported Haldane chain compounds, $A$Ni$_{2}$V$_{2}$O$_{8}$ where $A$ = Sr or Pb are of particular interest since they have substantial interchain interactions and single ion uniaxial anisotropy \cite{ZheludevPRB.62.8921,ZheludevPRB.64.134415}. The compound \srni\ is of present interest. The spin Hamiltonian for \srni\ can be defined as 
\begin{eqnarray}
H = J\sum_i S_i.S_{i+1} + J'\sum_i S_i.S_{i+2} +J_{\perp}\sum_{i,j} S_i.S_j \nonumber \\* 
+\sum_i D(S_i^z)^2 ,
\label{eq:one}
\end{eqnarray}
\noindent where the first and second terms are the nearest neighbor (NN) and next-nearest neighbor (NNN) intrachain interactions with exchange constants $J$, and $J'$, respectively. The third term is due to interchain interactions where $J_{\perp}$ is the effective interchain exchange constant. The fourth term is due to single-ion-anisotropy parameter $D$.

\srni\ was proposed to be situated close to the phase boundary between the ordered and spin-liquid states in the Sakai-Takahashi phase diagram \cite{ZheludevPRB.62.8921}. Based on initial powder inelastic neutron scattering (INS) studies, Zheludev {\it{et al.}} \cite{ZheludevPRB.62.8921,ZheludevPRB.64.134415} proposed an ordered AFM ground state for \srni\ unlike the iso-structural compound PbNi$_2$V$_2$O$_8$. However, a later nuclear magnetic resonance (NMR) study on powder samples of \srni\ by Pahari {\it{et al.}} \cite{PahariPRB.73.012407} found a nonmagnetic spin-liquid (singlet) ground state. Moreover, our earlier bulk properties and preliminary neutron scattering measurements on a single crystal confirmed that \srni\ has a non-magnetic spin-singlet ground state and gapped magnetic excitations \cite{BeraPRB.87.224423}. 

Our previous single crystal study also revealed that a magnetic field induced quantum phase transition occurs from the gapped spin-liquid state to a 3D AFM ordered state \cite{BeraPRB.87.224423}. The AFM ordered state is in contrast to the theoretical prediction of a field-induced ‘Luttinger liquid’ state for an isolated Haldane chain without anisotropy \cite{KonikPRB.66.144416}. Moreover, when comparing the gap values to the critical magnetic fields, the Boson model \cite{AffleckPRB.46.9002,FarutinJEPT.104.751} correctly estimates gap value, in contrast to the Fermion and perturbative models \cite{AffleckPRB.46.9002,GolinelliJPCM.5.7847}. In contrast, for most of the studied Haldane chain compounds, which have planar anisotropy, the Fermion model estimates the gap values more accurately \cite{Regnault.PRB.50.9174, Zheludev.PRB.68.134438}. In agreement with our finding, the gap values of the isostructural compound PbNi$_2$V$_2$O$_8$ which also has uniaxial anisotropy were also better described by the Bosonic model \cite{SmirnovPRB.77.100401}. It was predicted that the interchain interactions and the uniaxial single-ion anisotropy in \srni\ play a crucial role in determining the magnetic properties as well as in inducing the 3D AFM ordered state above the critical magnetic field ($H_c$). However, the nature of interchain interactions and anisotropy, as well as their role in the spin correlations are still unknown and are of present research interest.   

In this paper, we report the results of single crystal inelastic neutron scattering measurements on \srni\ and DFT calculations to investigate the role of complex interchain interactions and single-ion anisotropy on the Haldane chain. Neutron scattering measurements were performed over a wide reciprocal space region to probe several Brillouin zones as well as over a wide energy range to map out the dispersions fully. The strengths of intrachain and interchain interactions are estimated from the experimental excitation spectra by the Random phase approximation (RPA) method, and the exchange interaction geometry is proposed. The magnetic interactions are further investigated and verified by DFT based electronic band structure calculations. These calculations also provide insight into the complex superexchange paths which give rise to these interactions. Finally, the position of the \srni\ in the theoretical $D$-$J_\perp$ (Sakai-Takahashi) phase diagram has been proposed showing where it lies within the spin-liquid phase. 

\section {Methods}

Single crystals of \srni\ were grown in the Crystal Laboratory at Helmholtz Zentrum Berlin f{\"u}r  Materialen and Energie (HZB), Berlin, Germany. The starting material for crystal growth was prepared from high purity powders of SrCO$_3$ (99.994{\%}, Alfa Aesar, Puratronic), NiO (99.998{\%}, Alfa Aesar, Puratronic)  and V$_2$O$_5$ (99.99{\%} Alfa Aesar, Puratronic) by solid state reactions. Crystal  growth was performed in a four mirror type optical image furnace (Crystal Systems Corp., Japan) by the Traveling-solvent-floating-zone (TSFZ) technique. After growth the as-grown crystals were checked by x-ray and neutron Laue diffraction. Ground single crystals were also checked with x-ray powder diffraction (Brucker D8) for phase purity. Detail of the crystal growth and characterization will be reported in a separate paper \cite{Islam.unpublished}.

A powder neutron diffraction pattern of \srni\ was recorded at 2 K by using the high resolution powder diffractometer E9 ($\lambda$ = 1.7982~\AA), at HZB, Germany. The diffraction pattern was analyzed by the Rietveld method using the FULLPROF software \cite{Fullprof}.

Excitation spectra were measured by single crystal inelastic neutron scattering using the triple axis spectrometers with incident beams of thermal neutrons (PUMA at Heinz Maier-Leibnitz Zentrum (MLZ), Garching, Germany and IN8 at Institut Laue-Langevin (ILL), Grenoble, France) as well as cold neutrons (PANDA at MLZ, Garching, Germany). For PUMA and IN8, the measurements were performed with fixed final neutron wave vector $k_f$ = 2.662~$\angstrom^{-1}$ and a Pyrolytic Graphite filter on the scattered side was used to remove higher order neutrons. For PANDA, the final neutron wave vector was fixed to $k_f$ = 1.57~$\angstrom^{-1}$ and a cooled beryllium filter was used. Magnetic excitations were measured at base temperatures [at 3.1 and 3.5~K (using CCRs) for PANDA and PUMA, respectively, and at 1.5~K (using an orange cryostat) for IN8]. All measurements were performed on a large cylindrical single crystal of mass $\sim$ 2.5~g (diameter: $\sim$ 6~mm and length: $\sim$ 30~mm) which was mounted on an aluminum sample holder. 

Density functional theory (DFT) calculations were performed using the full-potential code \textsc{fplo9.07-41} \cite{dft:fplo}.  For the generalized gradient approximation (GGA) and GGA+$U$ calculations, we used the parameterization from Ref.~\onlinecite{dft:pbe96}.  All calculations were performed for the experimentally determined crystal structure at 2~K. Nonmagnetic GGA calculations were done on a 16$\times$16$\times$16 $k$-mesh (648 points in the irreducible wedge).  Wannier functions centered on Ni sites were computed according to the procedure described in Ref.~\onlinecite{dft:fplo_wf}.  For the local $x$ and $y$ axes we used Ni--O1 and Ni--O4 vectors, respectively.  Spin-polarized GGA+$U$ calculations were done using symmetry-reduced cells (sp.\ gr.\ $P1$) on a 2$\times$2$\times$2 $k$-mesh.  We used the on-site Coulomb repulsion of $U_d$\,=\,4.5\,eV and the Hund's exchange of $J_d$\,=\,1\,eV within the fully localized limit.

\section {RESULTS AND DISCUSSION}

\subsection{Crystal structure}

\begin{figure}
\includegraphics[trim=0cm 0cm 0cm 0cm, clip=true, width=80mm]{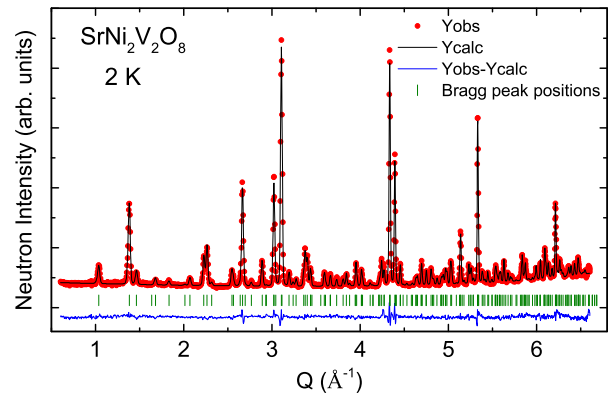} 
\caption{\label{Fig:ND2K}(Color online) The observed (filled circles) and calculated (solid black line) neutron diffraction patterns for \srni\ measured at 2~K. The difference between observed and calculated patterns is shown by the thin blue line at the bottom. The vertical bars give the allowed Bragg peak positions.}
\end{figure}

First we discuss briefly the crystal structure of \srni\ which is important to understand the exchange interaction geometry. Powder neutron diffraction at 2~K using the E9 diffractometer at HZB, Berlin, Germany, was used to determine the low temperature structural parameters. The parameters were used for the analysis of the low temperature magnetic excitation spectra. The Rietveld refined diffraction pattern is shown in Fig.~\ref{Fig:ND2K}. The analysis confirms that the compound crystallizes in the tetragonal space group $I$4$_1cd$  as found earlier at room temperature \cite{BeraPRB.86.024408}. The refined lattice parameters are $a$ = $b$ = 12.135(1)~\AA\ and $c$ = 8.3159(1)~\AA. The positions of Sr, Ni and O were refined during the analysis. The refined values of their fractional coordinates at 2~K are given in Table ~\ref{T:str2K}. Since neutrons are not sensitive to the V nuclear position, the respective atomic coordinates were adopted from Ref.~\onlinecite{dft:str} and were kept fixed during the refinement. 

\begin{table}
\caption{\label{T:str2K} 
The crystal structural parameters of \srni\ determined by neutron diffraction at 2~K. The space group is $I4_1cd$ (110), $a$~=~12.135(1)~\AA, $c$~=~8.3159(1)~\AA. Full occupancies for all the atomic sites were considered.}
\begin{ruledtabular}
\begin{tabular}{l c r r r r}
\multirow{2}{*}{Atom} & \multicolumn{1}{c}{Wyckhoff}  \\
& \multicolumn{1}{c}{position} &
  \multicolumn{1}{c}{$x/a$} & \multicolumn{1}{c}{$y/a$} & \multicolumn{1}{c}{$z/c$}
& \multicolumn{1}{c}{$B_{\text{iso}}$ (\r{A}$^2$)}  \\ \hline
Sr   &   $8a$ & 0         & 0          & $0.25$ & 0.87(1) \\
V\footnote{the values of the atomic coordinates and  $B_{iso}$ were adopted from Ref. \onlinecite{dft:str}}    &  $16b$ & 0.0802  & 0.2596  & 0.1485   & 0.22 \\
Ni   &  $16b$ & 0.3304(4) & 0.1678(3)  & 0.2793(1)     & 0.40(4) \\
O1   &  $16b$ & 0.1476(5) & 0.0019(10) & 0.5051(15)    & 0.49(1) \\
O2   &  $16b$ & 0.1601(9) & 0.3335(9)  & 0.0214(11)    & 0.77(3) \\
O3   &  $16b$ & 0.3196(7) & 0.1623(3)  & 0.0371(1)     & 0.26(3) \\
O4   &  $16b$ & 0.3326(6) & 0.0021(9)  & 0.3071(12)    & 0.41(0) \\
\end{tabular}
\end{ruledtabular}
\end{table}

\begin{figure}
\includegraphics[trim=0.5cm 0.5cm 5.2cm 0.5cm, clip=true, width=80mm]{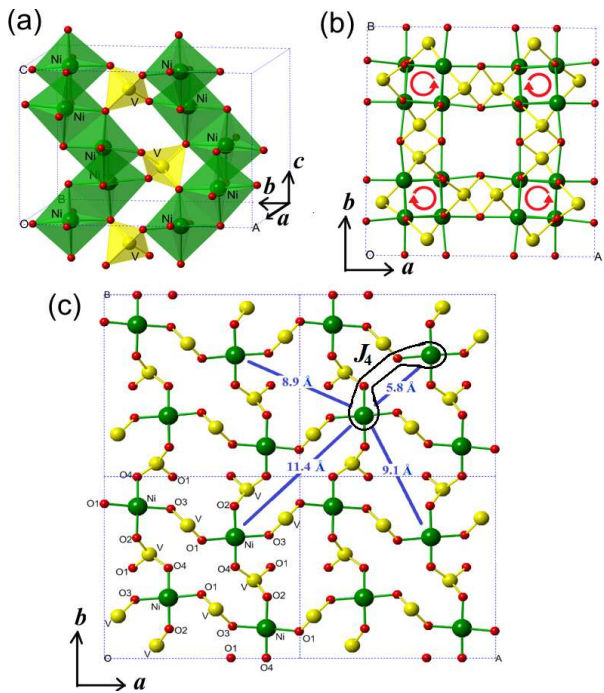} 
\caption{\label{Fig:Cryst_struc}(Color online) (a) The crystal structure of \srni. The Ni, V and O -ions are shown by green (large), yellow (medium) and red (small) spheres, respectively. The dotted  lines indicate the dimensions of the unit cell. For clarity only two chains (out of four chains per unit cell) are shown. (b) Projection of the crystal structure on to the $ab$ plane. Only the atoms and bonds are shown without polyhedra for clarity. Ni-O and V-O bonds are shown by green and yellow lines, respectively. Red arrows represent the direction of the rotation of the screw chains when propagating along the $c$ axis. Neighboring chains rotate in opposite directions. (c) Arrangement of the Ni, V, and O atoms within a given $ab$ plane (at $c$ = 0). The direct distances between Ni ions along the diagonal directions are shown. The shortest diagonal corresponds to the exchange interaction $J_4$ which occurs via Ni--O..O--Ni pathway as indicated.}
\end{figure}

The spin chains in \srni\ are formed by edge-shared NiO$_6$ (Ni$^{2+}$; 3$d^8$, $S$ = 1) octahedra along the $c$ axis [Fig.~\ref{Fig:Cryst_struc}(a)]. The screw chains are connected by nonmagnetic VO$_4$ (V$^{5+}$; 3$d^0$, $S$ = 0) tetrahedra. There are four screw chains per unit cell which are centered around (1/4, 1/4); (1/4, 3/4); (3/4, 1/4); and (3/4, 3/4) in the $ab$ plane [Fig.~\ref{Fig:Cryst_struc}(b)]. Each of these screw chains is 4-fold and contains four Ni ions along the $c$ axis within one unit cell. The two diagonal chains rotate clockwise while the other two chains rotate anti-clockwise when propagating along the $c$ axis. Furthermore, a phase shift of $c$/2 is present between diagonal chains [between (1/4, 1/4) and (3/4, 3/4), and between (1/4, 3/4) and (3/4, 1/4)]. This results in a $90^\circ$ rotation of all the bonds within an $ab$ plane [as shown in Fig.~\ref{Fig:Cryst_struc}(c)] with respect to the neighboring $ab$ plane along the $c$ axis. This peculiar crystal structure provides multiple interchain interactions between neighboring chains \cite{BeraPRB.86.024408}. The interchain interactions are not confined within the $ab$ planes but can also have an out-of-plane component. For instance, the interchain interactions are possible between two Ni$^{2+}$ ions (from two neighboring chains) which have an offset relative to each other of  $c$/4 and $c$/2 along the $c$ axis. In addition, next-nearest neighbor intrachain interactions, coupling two Ni ions separated by $c$/2, are also present due to the screw nature of the chains. 

\subsection{Inelastic neutron scattering}

\begin{figure*} 
\includegraphics[trim=0cm 0cm 0cm 0cm, clip=true, width=170mm]{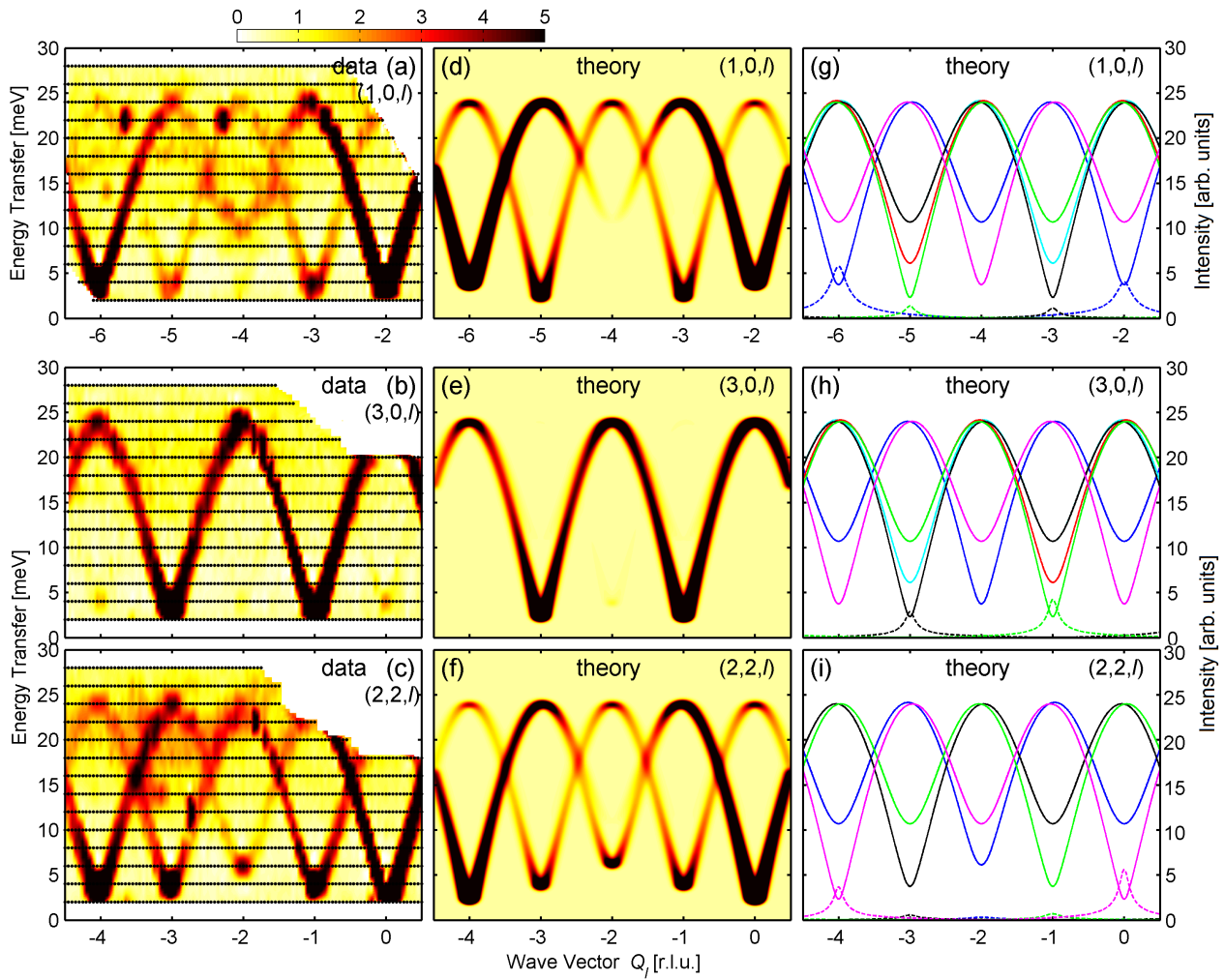}
\caption{\label{Fig:disp-chain}(Color online) The measured and calculated magnetic excitation spectra. Left panel [(a), (b), and (c)] shows the measured magnetic excitation spectra along (1,0,$l$), (3,0,$l$), and (2,2,$l$), respectively, at 3.5 K using the triple axis spectrometer PUMA. The excitation spectra were obtained by combining several constant energy scans with an interval of 2 meV. Black dots represent the points of measurements. The panel at the centre [(d) (e), and (f)] shows the simulated magnetic excitation spectra along (1,0,$l$), (3,0,$l$), and (2,2,$l$), respectively, as per the model discussed in the text and using the best fit parameters (Table~\ref {tab:interactions}). The calculated spectra were convoluted with a Gaussian function and corrected for the magnetic form factor. The intensities are in arbitrary units. r.l.u. stands for reciprocal lattice units. Some additional intensities (sharp and diffuse) in the measured spectra are due to the contributions from phonon scattering. The right panel [(g), (h), and (i)] shows the calculated dispersion relations (solid curves) and intensities (dashed curves at the bottom) of each of the six individual modes that appear due to the screw nature of the chains (see text for details). For simplicity dispersion curves are shown without considering the anisotropy induced splittings. Different colors (blue, red, green, magenta, black and cyan) are used to represent different modes.}
\end{figure*}

The excitation spectra [$S(q,\omega)$] along the chain direction, measured at 3.5~K using the PUMA spectrometer, show gapped magnetic excitations with strongly dispersive modes up to $\sim$ 23~meV [Figs.~\ref{Fig:disp-chain}(a), \ref{Fig:disp-chain}(b) and \ref{Fig:disp-chain}(c)]. Complicated variations in the intensity of the excitation modes are found. In addition to the intensity modulation of Haldane chains (as discussed in Fig.~\ref{Fig:Haldane}), additional complexity arises due to the screw nature of the spin-chains in \srni. As discussed in the previous section the screw chains have four fold periodicity along the chain axis and each of the magnetic ions is shifted from the center of the chain axis. This leads to step-4 periodicity in the transverse displacement of the magnetic sites from the chain axis. Exact analytical calculation shows that the screw chains result in six modes with different structure factors \cite{ZheludevPRB.62.8921}. All the individual modes have a periodicity of 4 r.l.u. along the chain direction (along the $Q_l$) since there are four Ni ions per unit cell along the chain axis. However, the modes are shifted in the reciprocal space (discussed later in detail). In a measurement along the $Q_l$ [such as along (1,0,$l$), (3,0,$l$), and (2,2,$l$) in Figs.~\ref{Fig:disp-chain}(a)--(c)], the individual modes appear with a shift of integer units with respected to each other [Figs.~\ref{Fig:disp-chain}(g)--(i)]. This leads to energy minima of one of the modes at each of the reciprocal lattice points (any integer combinations of $h$, $k$, and $l$) which are refered to as AFM zone centers [Figs.~\ref{Fig:disp-chain}(g)--(i)]. As the structure factor of individual modes is decided by the momentum transfer perpendicular to the chain axis (values of $h$ and $k$) (discussed later in detail) a substantial variation of relative intensity of the modes appears in the measured spectra [Fig.~\ref{Fig:disp-chain}(a)--(c)]. For example, in the case of (3,0,$l$) finite intensities are present only for the black [AFM zone center at (3,0,-3)] and the green [AFM zone center at (3,0,-1)] modes [Fig.~\ref{Fig:disp-chain}(h)]. Intensities for the other four modes (red, blue, cyan, and magenta) are zero over the given reciprocal space.

\begin{figure}
\includegraphics[trim=1.6cm 0.5cm 2.8cm 0.2cm, clip=true, width=86mm]{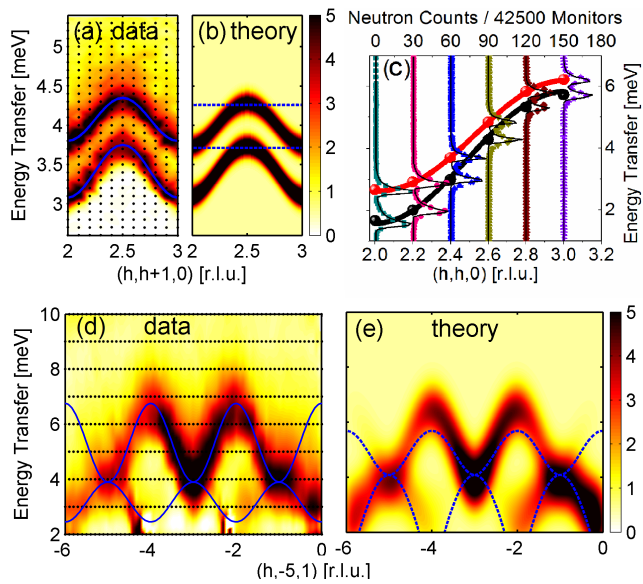}
\caption{\label{Fig:disp-perpchain}(Color online) The measured and simulated  magnetic excitation spectra perpendicular to the chain axis. (a) and (c) Measured excitation patterns using the PANDA spectrometer at 3~K, (d) using the IN8 spectrometer at 1.5~K, respectively. Black dots in (a) and (d)  represent the points of measurements. The solid blue curves in (a) and (d) are the fitted dispersion curves. (b) and (e) The simulated spectra, correspond to (a) and (d), as per the model discussed in the text using the best fit parameters (Table~\ref {tab:interactions}). For the comparison, simulated dispersion curves with $J_4$ = 0 (see text) are also shown in (b) and (e) by the dashed lines. The thin black lines through the measured points in (c) are the Gaussian fits to the data which are convoluted with instrumental resolution. The big solid circular points are the fitted peak positions and the thick lines are the fitted dispersion curves as per the model discussed in the text. }
\end{figure}

\begin{figure}
\includegraphics[trim=0.2cm 0.2cm 0.2cm 0.2cm, clip=true, width=86mm]{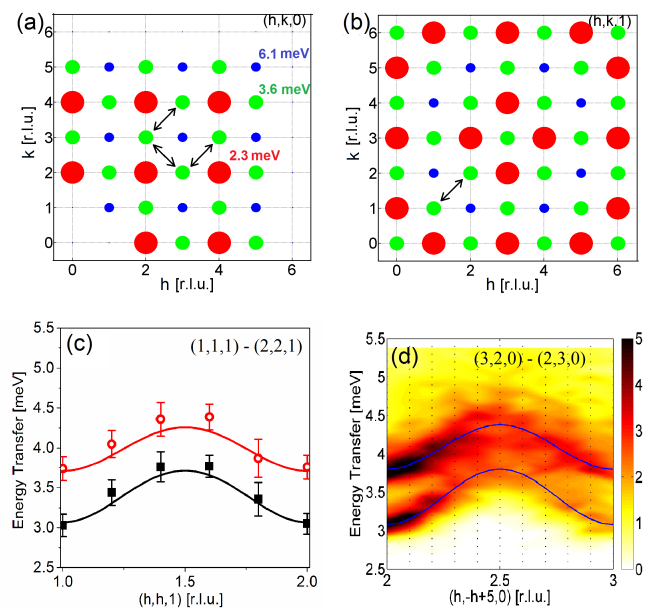}
\caption{\label{Fig:diagonal_exchange}(Color online) The measured energy minima patterns in the (a) ($hk0$) and (b) ($hk1$) reciprocal planes, respectively. The big (red), medium (green) and small (blue) points  correspond to the average values of lowest (2.3~meV), medium (3.6~meV) and highest (6.1~meV) energy minima, respectively. The arrows indicate the measured dispersion along the diagonal directions. (c) and (d) The measured dispersions between (1,1,1) and (2,2,1), and between (2,3,0) and (3,2,0) reciprocal points, respectively. The solid lines are the fitted dispersion curves.}
\end{figure}

Three different energy minima (sizes $\sim$ 2.3~meV, $\sim$ 3.6~meV and $\sim$ 6.1~meV)  are found at different AFM zone centers due to the complex interchain couplings. Furthermore, high resolution measurements using the cold neutron PANDA spectrometer [Figs.~\ref{Fig:disp-perpchain}(a) and \ref{Fig:disp-perpchain}(c)] confirm that each of the modes is split into two branches by anisotropy, giving the energy minima: (i) 1.57 $\pm$ 0.01 and 2.58 $\pm$ 0.01~meV, (ii) 3.07 $\pm$ 0.01 and 3.76 $\pm$ 0.01~meV, and (iii) 5.78 $\pm$ 0.01 and 6.18 $\pm$ 0.01~meV, respectively. The measured energy minima patterns for the AFM zone centers in two representative reciprocal planes $(hk0)$ and $(hk1)$ are shown in Figs.~\ref{Fig:diagonal_exchange}(a) and \ref{Fig:diagonal_exchange}(b), respectively. For simplicity, only the average values of the anisotropy split gaps are presented. Complex patterns arise due to several competing interchain interactions (discussed later) and the screw nature of the chains. It is evident that all the energy minima values along some diagonal directions are the same (intermediate value of 3.6~meV) for particular combination of $h$ and $k$ values i.e., even-odd for the $l=0$ and both even-even and odd-odd for the $l=1$. It may be worth to mention that the gap minima [red points in Figs.~\ref{Fig:diagonal_exchange}(a) and \ref{Fig:diagonal_exchange}(b)] become magnetic Bragg peaks in a magnetic field ($H > H_c$).

The presence of dispersions [Figs.~\ref{Fig:disp-perpchain} and \ref{Fig:diagonal_exchange}] perpendicular to the chain direction confirms that there are finite interchain interactions. The maximum bandwidth of the perpendicular dispersions ($\sim 4.0 $ meV) is about 20 $\%$ of the dispersion along the chain-direction. For the dispersion along ($h$,-5,1) [Figs.~\ref{Fig:disp-perpchain}(d)], two modes (anti-modes) are present with the same periodicity (2 r.l.u.) in which one mode disperses between $\sim$~2.5--3.9 meV (lower energy mode) and the other mode disperses between $\sim$~3.9--6.6 meV (higher energy mode). It may be noted that the bandwidth of the higher energy mode ($\sim$~2.7 meV) is larger than that of the lower energy mode ($\sim$~1.4 meV) which is found to be due to the competing interchain interactions (discussed later in detail). In addition, dispersions are also found along the diagonal directions between two neighboring reciprocal points having intermediate energy minima values (mean value of 3.6~meV) as shown in Fig.~\ref{Fig:disp-perpchain}(a), Fig.~\ref{Fig:diagonal_exchange}(c), and Fig.~\ref{Fig:diagonal_exchange}(d). These dispersions indicate the presence of interchain interactions along the diagonal directions as found from the detailed analysis (discussed later). 

\begin{figure} [t]
\includegraphics[width=80mm]{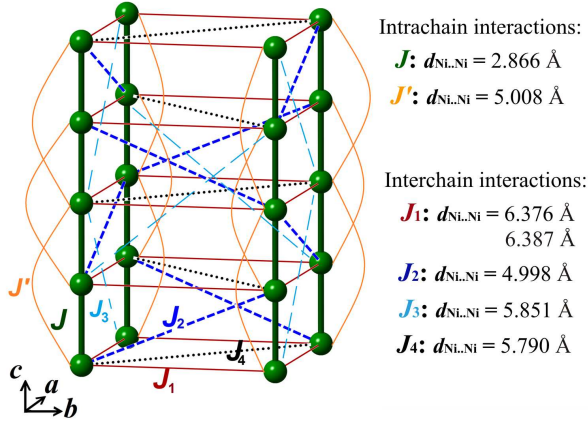}
\caption{\label{Fig:exchange}(Color online) The magnetic interaction topology of \srni\ (mapped onto equivalent straight chains). $J$: nearest neighbor intrachain interaction, $J^{'}$: next nearest neighbor intrachain interaction, $J_1$: between two Ni ions from neighboring chains having same $c$ values (in the $ab$ plane), $J_2$: between two Ni ions from neighboring chains having offset of $c$/4, $J_3$: between Ni ions from neighboring chains having offset of $c$/2, and $J_4$: between Ni ions from diagonal chains within the $ab$ plane. The Ni..Ni direct distances in the real crystal structure for corresponding interactions are also given.}
\end{figure}

To explain the observed features in the excitation spectra and to investigate the interactions and anisotropy, a model of the magnetic dynamical structure factor $S$($q$, $\omega$) (Eq. \ref{eq:Struc_fact}) was derived by the RPA method for the Hamiltonian (Eq. \ref{eq:one}) according to the method described by Zheludev {\it et. al.} in Ref. \onlinecite{ZheludevPRB.62.8921} and Ref. \onlinecite{ZheludevPRB.64.134415}. The  $S$($q$, $\omega$) for \srni\ was calculated in two steps; (i) first the calculation of $S^{'}$($q$,~$\omega$) for an equivalent Bravais spin-lattice of straight chains with an appropriate interaction topology [Fig.~\ref{Fig:exchange}] and then (ii) the adaptation of the results for the more complex real crystal structure of \srni\ having screw chains with a fourfold periodicity  \cite{ZheludevPRB.62.8921}. 

The dynamical structure factors of a straight chain (in the single mode approximation) along the longitudinal ($S_{\parallel}^{'}$) and transverse ($S_{\perp}^{'}$) directions are given as  
\begin{eqnarray}
S_{\parallel, \perp}^{'}(q, \omega) = \frac{1 - cos (q)}{2}
\times \frac{Z \nu}{\hbar \omega_{\parallel, \perp} (q)} \delta[\hbar \omega - \hbar \omega_{\parallel, \perp} (q)]  \nonumber  \\*
\label{eq:struct_fact_SC}
\end{eqnarray}
\noindent where $\nu$ (= 2.49$J$) is the spin-wave velocity, and $Z$ = 1.26 \cite{ZheludevPRB.62.8921} and $q$ is the reduced momentum transfer along
the chain direction. For \srni, $q$ = $Q_c$/4 = $l$/4 where $Q$ (= $ha^*+kb^*+lc^*$) is the actual wave vector transfer, since there are four equivalent Ni$^{2+}$ ions per unit cell along the chain direction ($c$ axis).

The dispersion relations for a coupled Haldane chain, like \srni, can be written (within the RPA) as
\begin{eqnarray}
[\hbar \omega_{\parallel, \perp} (q)]^2 = \Delta_{\parallel, \perp}^2+ \nu^2 sin^2 (\pi l/2) + \alpha^2 cos^2 (\pi l/4) \nonumber \\ 
+ \frac{Z \nu}{2} J^{'}cos(\pi l) [1 - cos(\pi l/2)] + \frac{Z \nu}{2} \mathcal{J} (Q) [1 - cos(\pi l/2)] \nonumber  \\*
\label{eq:disp}
\end{eqnarray}
where $\Delta_{\parallel, \perp}$ are the longitudinal and transverse Haldane gaps, respectively, for noninteracting chains. Two energy gaps appear due to the zero-field splitting of  the triplet states by single-ion anisotropy \cite{RegnaultJPCM.5.L677}. The relations between the energy gaps and single-ion anisotropy were proposed as \cite{GolinelliJPCM.5.7847,GolinelliPRB.46.10854} 
\begin{eqnarray}
\Delta_{\perp} = \Delta_0 - 0.57D  \text{			and		} \Delta_{\parallel} = \Delta_0 + 1.41D ,
\label{eq:aniso}
\end{eqnarray}
where $\Delta_0$ is the intrinsic Haldane gap ($\approx$ 0.41$J$) \cite{WhitePRB.77.134437} in the absence of single-ion anisotropy. The 2nd and 3rd terms give the intrachain dispersions. The $\alpha$ (=1.1$J$) in Eq. \ref{eq:disp} characterizes the asymmetry of the dispersion relation \cite{ZheludevPRB.64.134415, ZaliznyakPRL.87.017202}. The asymmetry arises due to the fact that the translational symmetry is not broken in the Haldane ground state \cite{TakahashiPRL.62.2313}. The fourth term is due to the contribution from the NNN intra-chain exchange interaction $J^{'}$ which is possible due to the screw chain crystal structure of edge sharing NiO$_6$ tetrahedra [Fig.~\ref{Fig:Cryst_struc}]. The $\mathcal{J} (Q)$ in Eq.~\ref{eq:disp} is the combined Fourier transform of all interchain interactions between the magnetic sites in a crystallographic unit cell (Bravais spin-lattice) [earlier referred as $J_{\perp}$ in Eq.~\ref{eq:one}].

Now we consider the interchain interactions for the straight-chain spin-lattice. Only the interactions which are possible in the real screw-chain crystal structure of \srni\  are considered (Fig.~\ref{Fig:exchange}). The interactions are (i) $J_1$: between two Ni ions from neighboring chains having the same $c$ values (in the $ab$ plane) along the crystallographic $a/b$ directions [$d_{\text{Ni..Ni}} = 6.378/6.387~\angstrom$ in the real crystal structure and the coordination number ($z$) is four], (ii) $J_2$: between two Ni ions from neighboring chains offset by $c$/4 [$d_{\text{Ni..Ni}} = 4.998~\angstrom$ in the real crystal structure and the coordination number $z$ = 2], (iii) $J_3$: between two Ni ions from neighboring chains having an offset of $c$/2 [$d_{\text{Ni..Ni}} = 5.851~\angstrom$ in the real crystal structure and the coordination number $z$ = 2], and (iv) $J_4$: between two Ni ions from diagonal chains within the $ab$ plane [$d_{\text{Ni..Ni}} = 5.79~\angstrom$ in the real crystal structure and the coordination number $z$ = 1]. Combining all these contributions, $\mathcal{J} (Q)$ becomes
\begin{eqnarray}
\mathcal{J} (Q) = &(2J_1 + J_2 cos(\pi l/2)+ J_3 cos(\pi l))[cos(\pi h)+cos(\pi k)] \nonumber \\*
&\hspace{5mm}+(J_4/2)cos(\pi h) cos(\pi k) 
\label{eq:inter_int}
\end{eqnarray}

Finally, the dynamical structure factors $S$($q$, $\omega$) for the real crystal structure of \srni\ are given by [Ref. \onlinecite{ZheludevPRB.62.8921}]
\begin{eqnarray}
&& S_{\parallel, \perp} (q, \omega) = cos^2\psi_1 cos^2\psi_2 S_{\parallel, \perp}^{'}(h,k,l)+ \frac{cos^2\psi_1 sin^2 \psi_2}{2}  \nonumber \\* 
&&  \times [S_{\parallel, \perp}^{'} (h+1,k,l+1)+S_{\parallel, \perp}^{'}(h+1,k,l+3)]\nonumber \\* 
&& + \frac{sin^2\psi_1 cos^2 \psi_2}{2} 
 \times [S_{\parallel, \perp}^{'} (h,k+1,l+1)+S_{\parallel, \perp}^{'}(h,k+1,l\nonumber \\*
&& +3)]+sin^2\psi_1 sin^2\psi_2 S_{\parallel, \perp}^{'}(h+1,k+1,l+2) ,
\label{eq:Struc_fact}
\end{eqnarray}

\noindent where 
\begin{eqnarray}
\psi_1 =\frac{2 \pi d}{a}h  \text{			and		} \psi_2 =\frac{2 \pi d}{a}k	 
\label{eq:phase}
\end{eqnarray}
are the 3D structure factors that arise due to the screw nature of the spin chains and $d$ ($\sim$ 0.8$a$) is the offset of each Ni$^{2+}$ ion along the $a$ or $b$ axis with respect to the central axis of the corresponding screw chains [(1/4, 1/4) or (1/4, 3/4) or (3/4, 1/4) or (3/4, 3/4)]. The dynamical structure factors $S$($q$, $\omega$) were simulated by using Eq. \ref{eq:Struc_fact} and compared to the measured  spectra. By adjusting the values of the interactions a good solution could be found which is given in Table \ref{tab:interactions} and the corresponding simulated excitation patterns are depicted in Fig.~\ref{Fig:disp-chain} and Fig.~\ref{Fig:disp-perpchain} alongside the measured patterns. 

\begin{table}
\caption{\label{tab:interactions}The fitted values of the exchange interactions from neutron scattering. All values are in meV units.} 
\begin{ruledtabular}
\begin{tabular}{ccccc}
 Interac-& $J$ & $J^{'}$ & ($2J_1-J_2+J_3$) & $J_4$ \\
tion& & & & \\
\hline
Value & 8.70$\pm$0.05 & 0.149$\pm$0.004 & 0.286$\pm$0.006 &0.32$\pm$0.01 \\
$D$& $-0.32$$\pm$0.01 &&&\\
\end{tabular}
\end{ruledtabular}
\end{table}

The analysis of the neutron scattering data reveals that the strongest interaction is the nearest neighbor intrachain AFM interaction $J$ = 8.7 meV. All other interactions are quite weak compared to $J$, although, comparable in strength to each other. Note that in the proposed model it is not possible to determine the individual values of $J_1$, $J_2$ and $J_3$ since these interactions result in virtually identical dispersion relations. Only the linear combination of these parameters, ($2J_1 - J_2 + J_3$) (since the coordination numbers for $J_1$, $J_2$ and $J_3$ are $z$ = 4, $z$ = 2, and $z$ = 2, respectively) can be determined and is found to be AFM in overall. The minus sign of $J_2$ expresses the fact that it competes with $J_1$ and $J_3$ if it has the same sign (e.g., AFM) but reinforces if it has opposite sign (e.g., FM). The diagonal interchain interaction $J_4$ is AFM and frustrated. $J_4$ strongly influences the dispersion relation perpendicular to the chain direction, it (i) makes the band widths asymmetric [Fig.~\ref{Fig:disp-perpchain}(d)], (ii) introduces dispersion along the diagonal directions [Fig.~\ref{Fig:disp-perpchain}(a) and Figs.~\ref{Fig:diagonal_exchange}(c)-(d)], as well as (iii) reduces the overall gap values. The NNN intrachain exchange interaction $J^{'}$ is also AFM and frustrated which increases the gap values. The resulting effect of all the interchain interactions leads to the three different energy minima $\sim$~2.3, $\sim$~3.6, and $\sim$~6.1~meV [Figs.~\ref{Fig:disp-chain}, \ref{Fig:disp-perpchain}, and \ref{Fig:diagonal_exchange}] at the AFM zone centers as compared to the intrinsic Haldane gap of ($\Delta_0 = 0.41J$)=3.57~meV. The fitted value of the uniaxial single-ion anisotropy parameter $D$ is found to be $-0.32$~$\pm$~0.01~meV. The value of the anisotropy parameter $D$ is in good agreement with the value $D$ = -0.29~meV obtained  from the electron spin resonance measurements \cite{WangPRB.87.104405}.

\begin{figure}
\includegraphics[width=75mm]{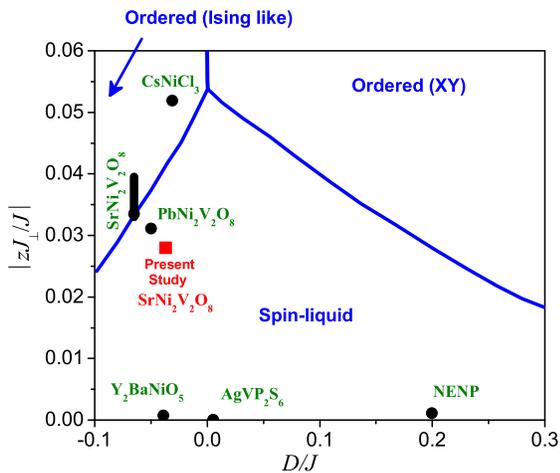}
\caption{\label{Fig:phase_dia}(Color online) The $D$-$J_{\perp}$ phase diagram for the weakly coupled Haldane chains (Ref. \onlinecite{SakaiPRB.42.4537}). The position of \srni\ obtained from the present study is shown by red square and compared with previous report (Ref.~\onlinecite{ZheludevPRB.62.8921}). Positions of the other experimentally studied Haldane chain compounds are also shown as per Ref. ~\onlinecite{ZheludevPRB.62.8921}.}
\end{figure}

In \srni, the values of all the interactions are such that the mean value (3.56~meV) of the intermediate energy minima (3.07 and 3.76~meV) is equal to the theoretically predicted intrinsic Haldane gap value of (0.41$J$ = 0.41 $\times$ 8.7~meV) $\sim$ 3.57~meV \cite{WhitePRB.77.134437}. The first availability of good quality large single crystal has enabled us to perform such detailed investigations which give accurate values of exchange interactions as well as single-ion anisotropy. The present investigation estimates the values $zJ_{\perp}/J$ = 0.028 and $D/J$ = $-0.37$~meV, respectively, and allows us to correctly position \srni\ in the theoretical $D$--$J_{\perp}$ phase diagram \cite{SakaiPRB.42.4537}, showing where it lies within the spin-liquid phase [Fig.~\ref{Fig:phase_dia}]. 

Now we focus on the diagonal interchain interaction $J_4$ which is of special interest. The presence of a diagonal interaction is strongly evident from the experimentally observed dispersions along the diagonal directions within the ($hk$) planes between two reciprocal lattice points having intermediate values (3.07~meV and 3.76~meV). For example, dispersions between (i) (2,3,0) and (3,4,0) [Fig.~\ref{Fig:disp-perpchain}(a)], (ii) (1,1,1) and (2,2,1) [Fig.~\ref{Fig:diagonal_exchange}(c)], as well as (iii) (2,3,0) and (3,2,0) [Fig.~\ref{Fig:diagonal_exchange}(d)]. The simulated curves without considering the diagonal interchain interaction (i.e., $J_4$~=~0) reveal no dispersion between these reciprocal lattice points. Such a simulated curve between the (2,3,0) and (3,4,0) reciprocal lattice points is shown by the dotted lines in Fig.~\ref{Fig:disp-perpchain}(b). The presence of frustrated AFM $J_4$ is further evident from the fact that the bandwidth of the higher energy mode is larger than that of the lower energy mode for ($h$,-5,1) [Fig. \ref{Fig:disp-perpchain}(d)]. Here again, the patterns simulated with $J_4$~=~0 do not match with the observed excitation patterns. In contrast to the data, the simulation with $J_4$ = 0 results in a larger bandwidth of the lower energy mode than that of the higher energy mode as shown by the dotted curves in Fig.~\ref{Fig:disp-perpchain}(e).

\begin{figure} 
\includegraphics[trim=0.5cm 0.5cm 0.5cm 0.5cm, clip=true, width=65mm]{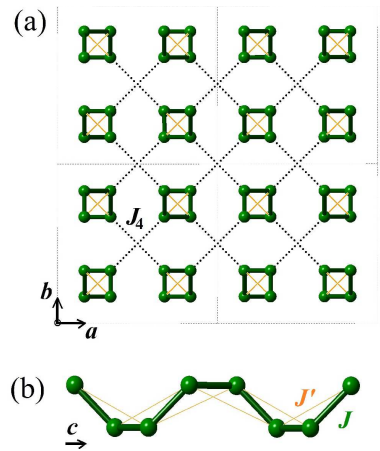}
\caption{\label{Fig:dia_path}(Color online) (a) A schematic representation of the diagonal exchange in \srni\ by a combination of inplane diagonal interchain interaction $J_4$ (dotted black line) and the NNN intrachain interaction $J^{'}$ (solid yellow line). (b) The NN and NNN  intrachain interactions $J$ and $J^{'}$, respectively, in a single screw chain.}
\end{figure}

Although the presence of diagonal interactions is confirmed from the experimental data, the nature of the exchange pathways is unclear for the real crystal structure of \srni. In the real crystal, the direct distances between Ni-ions along the diagonal directions are large $\sim$ 5.8 and 11.4, 8.9 and 9.1~\AA\ [Fig.~\ref{Fig:Cryst_struc}(c)] which are far beyond the limit of any possible direct exchange interactions. Moreover, along the diagonal directions, the Ni ions are not connected directly by any obvious pathways (constructed from Ni--O and V--O bonds) [Fig.~\ref{Fig:Cryst_struc}(c)] that can provide superexchange interactions. A superexchange interaction ($J_4$) is only possible between the Ni ions having shortest distance (5.8~\AA) via nontrivial Ni--O..O--Ni pathways [Fig.~\ref{Fig:Cryst_struc}(c)]. This is confirmed from the DFT calculations (discussed later).

The continuous propagation of this single $J_4$ interaction along a diagonal is not possible in \srni. If two Ni ions are separated along the diagonal by 5.8~\AA\ and coupled by $J_4$, the next Ni ion in the same diagonal direction will be situated far away (11.4~\AA) [Fig.~\ref{Fig:Cryst_struc}(c)]. Thus there is no possible continuous exchange path along the inplane diagonal direction (1,1,0). Nevertheless, an effective diagonal interaction along the body diagonal (1,1,1) direction may be plausible by a combination of $J_4$ and the NNN intrachain interaction ($J^{'}$). Such an interaction scheme is shown in Fig.~\ref{Fig:dia_path}. As mentioned in the crystal structure section all the bonds within a $ab$ plane rotate by 90 degrees with respect to the neighboring $ab$ planes along the $c$ axis due to the 4$_1$ glide symmetry of the space group $I4_1cd$. Therefore, the shortest diagonal distances (5.8~\AA) which are parallel to each other are shifted by $c$/2 along the $c$ axis. These two paths can be connected to each other by the NNN intrachain interaction ($J^{'}$) which occurs between two Ni ions shifted by $c$/2 along the $c$ axis within a given chain. Thus, a combination of $J_4$ and $J^{'}$ results in an effective diagonal interaction along the (1,1,1) direction.

\subsection{DFT calculations}

To get further insight in the microscopic exchange interaction model, we employed DFT calculations. As discussed in the previous section, the INS study can estimate the sums of different interchain couplings, yet in some cases their individual values remain unknown. The purpose of our DFT calculations is twofold. In DFT calculations, we can directly access all individual exchange couplings, our primary goal is to verify the exchange interactions provided by INS, to estimate their individual values, as well as supplement our microscopic magnetic model with the missing terms.  Second, we are interested in the mechanism of the magnetic exchange, i.e.\ which orbitals (including those on nonmagnetic sites) have a sizable contribution to the magnetic exchange.

\begin{figure}
\includegraphics[width=8.6cm]{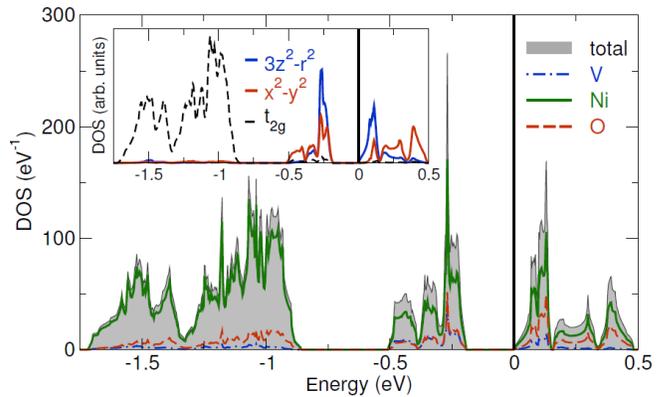}
\caption{\label{Fig:dos}
(Color online) The valence band of \srni\ as calculated using GGA.  Total and
atom-resolved density of states (DOS) are shown.  The Fermi level is at zero
energy.  Inset: the orbital-resolved DOS for the Ni $3d$ states.}
\end{figure}

Before we proceed to the DFT results, several remarks on the electronic structure of Ni$^{2+}$ systems should be made.  In \srni, the magnetic Ni atoms have a distorted octahedral environment. Hence, the electrical crystal field splits the $d$ orbitals into two manifolds: $t_{2g}$ comprising $xy$, $xz$, and $yz$ orbitals, and $e_g$, comprising $x^2-y^2$ and $3z^2-r^2$ orbitals. The $d^8$ electronic configuration of Ni$^{2+}$ implies that the lower-lying $t_{2g}$ orbitals are fully filled, while the $e_g$ orbitals are half-filled (one electron per orbital).  Since the number of electrons is even, we expect a gap in the electronic excitation spectrum (band insulator). This intuitive picture is fully supported by our non-magnetic GGA calculations that yield a valence band dominated by Ni states split into two manifolds and a gap of $\sim$0.2~eV (Fig.~\ref{Fig:dos}).  Such a small band gap renders \srni\ as a semiconductor, which is at odds with the yellow color of the crystals, indicating the importance of strong electronic correlations that are severely underestimated in the GGA. We should also note that the valence band is dominated by the Ni states, yet there is a substantial contribution of O and V states especially in the $e_g$ channel (Fig.~\ref{Fig:dos}).  Both O and V states play a crucial role for the intra-chain $J$ exchange, as will be demonstrated below.  

\begin{figure}
\includegraphics[width=8.6cm]{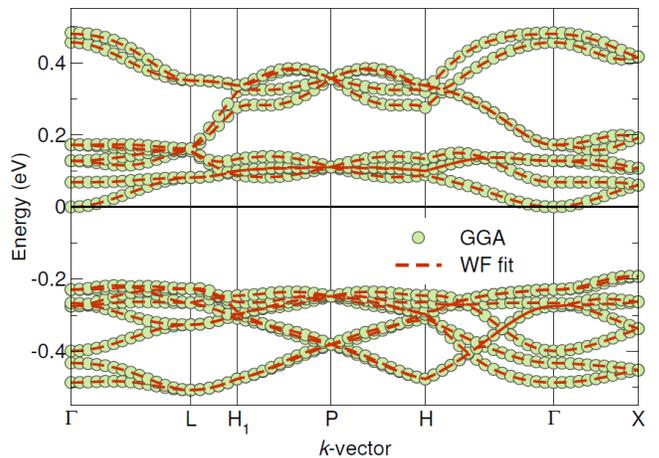}
\caption{\label{Fig:band}
(Color online) The GGA band structure of the $e_g$ states (circles) and the fit
using Ni $3d_{x^2-y^2}$ and $3d_{3z^2-r^2}$-centered Wannier functions (dashed
lines).  The Fermi level is at zero energy.}
\end{figure}

\begin{table*} 
\caption{\label{Tab:dft} 
Leading transfer $t_{ij}$ (in meV) and exchange $J_{ij}$ (in meV) integrals in
\srni. The paths are indicated in Fig.~{\ref{Fig:exchange}}.
Interatomic Ni..Ni distances given in \r{A} correspond to the 2~K structure (Table~\ref{T:str2K}).
Transfer integrals are provided for both magnetically active orbitals,
$3z^2-r^2$ and $x^2-y^2$.  Note that due to the lack of inversion symmetry
$t_{x^2-y^2\leftrightarrow{}x^2-y^2}\neq{}t_{x^2-y^2\leftrightarrow{}3z^2-r^2}$.
Terms with the absolute value larger than 5~meV are shown.  The second last two
columns show the GGA+$U$ value for the magnetic exchange and the respective
experimental values from INS.}
\begin{ruledtabular}
\begin{tabular}{l r r r r r r c c}
path &
$d_{\text{Ni..Ni}}$ &
$t_{3z^2-r^2\leftrightarrow{}3z^2-r^2}$ &
$t_{x^2-y^2\leftrightarrow{}x^2-y^2}$ &
$t_{x^2-y^2\leftrightarrow{}3z^2-r^2}$ &
$t_{x^2-y^2\leftrightarrow{}x^2-y^2}$ &
$J$ (GGA+$U$) & $J$ (INS) & interactions \\ \hline
$X$   & 2.866 &$-94$& $-111$&      84&    216 &   8.43  & 8.7 & $J$\\
$X'$  & 5.008 &   31&     35&   $-34$&     10 &   0.21  & 0.15 & $J^{'}$\\
\multirow{2}{*}{$X_1$} &
        6.376 &   12&       &        &   $-6$ & $-0.11$  & \multirow{4}{*}{$[(2J_1-J_2+J_3)$=0.29]} &\multirow{2}{*}{$J_1$}\\
      & 6.387 &   14&       &        &   $-6$ & $-0.06$  & \\
$X_2$ & 4.998 & $-8$&     20&        &      7 & $-0.28$  & & $J_2$\\
$X_3$ & 5.851 &    5&  $-21$&      10&      5 & $-0.08$ & & $J_3$\\
$X_4$ & 5.790 &$-13$&       &        &     45 &   0.16  & 0.32 & $J_4$\\
\end{tabular}
\end{ruledtabular}
\end{table*}

The primitive unit cell of \srni, whose volume is twice smaller than the conventional body-centered tetragonal cell, contains 8 magnetic Ni atoms, i.e.\ we expect 8$\times$5~=~40 bands with dominant Ni $d$ character.  Since the magnetism pertains to low-energy excitations, we restrict our analysis to the $e_g$ states (Fig.~\ref{Fig:band}) that lie in the vicinity of the Fermi energy. These 16 bands are projected onto a real-space basis of Ni-based Wannier functions, while the resulting overlap integrals $t_{ij}$ parameterize an
effective two-orbital ($x^2-y^2$ and $3z^2-r^2$) tight-binding model.  Going back to the $k$-space, the tight-binding Hamiltonian yields excellent agreement with the GGA bands (Fig.~\ref{Fig:band}).

\begin{figure} 
\includegraphics[width=7.0cm]{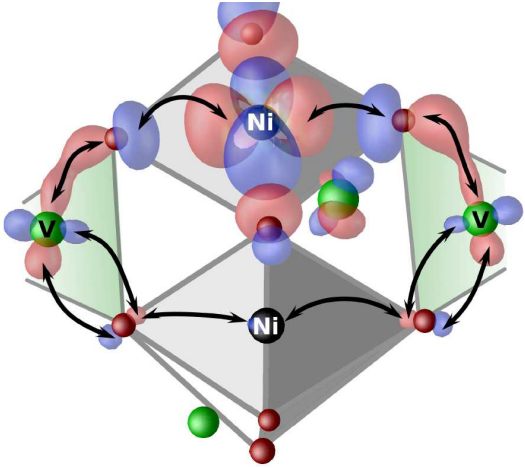}
\caption{\label{Fig:WF}
(Color online) 
The Wannier function centered at the Ni $x^2-y^2$ orbital.  The dominant
exchange coupling $J$ is realized by superexchange along the two inequivalent
Ni--O--V--O--Ni paths indicated by arrows. 
} \end{figure}

For each pair of Ni atoms, electron transfer can involve (i) two $x^2-y^2$ orbitals, (ii) two $3z^2-r^2$ orbitals, or two orbitals of different character. Since \srni\ lacks inversion symmetry, the $x^2-y^2\rightarrow{}3z^2-r^2$ and $3z^2-r^2\rightarrow{}x^2-y^2$ couplings are not equivalent, hence there are four $t_{ij}$ terms for each pair of Ni atoms $i$ and $j$. The resulting transfer integrals $t_{ij}$ are provided in Table~\ref{Tab:dft}. All the leading terms correspond to the NN intrachain coupling ($J$), and the $x^2-y^2$ contribution to this coupling is by far the largest. The AFM $J$ ($\sim$ 8.7~meV [Table~\ref{tab:interactions}]) is surprisingly strong for an almost $90^\circ$ Ni--O--Ni superexchange pathways between two intrachain NN Ni ions. According to the Goodenough-Kanamori rules \cite{Goodenough.PR.100.564,Kanamori.JPCS.10.87}, a ferromagnetic interaction is preferable for $90^\circ$ superexchange pathways. To pinpoint the cause of this dominance, we plot one of the Ni $d_{x^2-y^2}$-based Wannier function in Fig.~\ref{Fig:WF}. By examining the nonlocal contributions, we find that the bridging O atom in the Ni--O--Ni intrachain path only moderately contributes to the electron transfer, while the main contribution is facilitated by the V $d$-orbitals and follows the nontrivial Ni--O--V--O--Ni superexchange pathways.

Long-range superexchange paths involving empty $d$ orbitals are quite common in V$^{5+}$ systems.  For instance, Cu--O--V--O--Cu hopping gives rise to a sizable exchange of 330~K in the spin-chain compound AgCuVO$_4$ \cite{dft:moeller2009}. A similar hopping process in $\beta$-Cu$_2$V$_2$O$_7$ leads to the honeycomb-lattice magnetism \cite{dft:tsirlin2010}, which is not observed in isostructural $\alpha$-Cu$_2$P$_2$O$_7$ (coupled dimers, Ref.~\onlinecite{dft:janson2011}) and $\alpha$-Cu$_2$As$_2$O$_7$ (alternating chains, Ref.~\onlinecite{dft:arango2011}). Hence, we can expect that the replacement of the $d$-element V in \srni\ with a $p$ element, such as P and As, would also drastically affect the magnetic coupling regime.

After establishing the mechanism of the NN intrachain couplings $J$, we turn to the much smaller interchain terms as well as the NNN intrachain term. These transfer integral terms are weak and amount to several meV at most, and this extremely low magnetic energy scale severely impedes the accuracy of DFT results. As follows from Table~\ref{Tab:dft}, several interchain couplings in \srni\ have similar strengths. To estimate the numerical values of these small terms, we employ spin-polarized GGA+$U$ calculations for magnetic supercells and map the resulting total energies onto a classical Heisenberg model. The main drawback of this method is the dependence of the resulting exchange integrals on the $U_d$ value used in GGA+$U$. To eliminate this problem, we investigated a range of $U_d$ values and evaluated the optimal $U_d$ value of 4.5~eV, which reproduces the experimental value for $J$ (8.7~meV) within the error bars.

The resulting interchain exchange couplings are given in Table~\ref{Tab:dft}. A reasonably good agreement between the exchange interactions from the GGA+$U$ and INS is found. The possible intra and interchain interactions (for any transfer integral $t_{ij} \ge$ 5~meV) agree well with the INS results and, hence, further long-range couplings in \srni\ can be neglected. In agreement with the INS results, the DFT calculations reveal the presence of the AFM NNN intrachain interaction $J^{'}$. The DFT calculations provide the individual values of $J_1$, $J_2$, and $J_3$ which are indistinguishable in the INS. In agreement with the INS, the overall sign of ($2J_1 - J_2 + J_3$) is found to be antiferromagnetic. The DFT calculations further reveal that all the $J_1$, $J_2$, and $J_3$ are ferromagnetic and the $J_2$ is the strongest interaction among them. The presence of the antiferromagnetic $J_4$ is also confirmed between two Ni ions having the shortest distance of 5.8~\AA\ along the diagonal direction within a given $ab$ plane. 

The exchange mechanism for the diagonal interaction $J_4$ was also revealed by the Ni $d_{x^2-y^2}$-based Wannier function. By examining the nonlocal contributions, we find that the mechanism for $J_4$ is superexchange along Ni--O..O--Ni paths. It is quite efficient, because (i) the magnetically
active $x^2-y^2$ orbitals of the respective Ni atoms lie in the same plane and (ii) the direct O..O distance (2.84 \AA) is short enough to facilitate overlapping of O 2$p$ states. This is very similar to the situation in the edge-sharing chain cuprate LiVCuO$_4$ \cite{Nishimoto.EPL.98.37007}, where the same mechanism gives rise to next-nearest-neighbor
exchange interactions.   \\

\section {Summary and conclusions}
In summary, our detailed single crystal inelastic neutron scattering study reveals that the ground state of the Haldane chain compound \srni\ is a non-magnetic singlet, however, the low energy excitation spectra are significantly modified by the complex interchain interactions and single-ion anisotropy. The effective energy gap (1.57~meV) is reduced from the intrinsic Haldane gap value of 3.57~meV, and the system is pushed close to the phase boundary to long-range order. Comprehensive experimental data along with DFT calculations reveal the important interactions, and estimate their strengths as well as the size of the single-ion anisotropy that allows \srni\ to be accurately positioned in the theoretical $D$--$J_{\perp}$ phase diagram. The position of \srni\ within the spin-liquid phase is shown and compared with previous results. The presence of multiple competing inter-chain interactions is confirmed by both INS measurements and DFT calculations. The interchain interactions are quite weak (in the order of one tenth of meV) as compared to the strongest nearest neighbor intrachain interaction $J$ = 8.7~meV. The DFT analysis reveals the mechanism of the exchange interactions, for example, (i) the strongest intrachain exchange $J$ occurs via nontrivial Ni--O--V--O--Ni paths involving the empty $d$ orbitals of the V ion and (ii) the diagonal interaction $J_4$ occurs via Ni--O..O--Ni paths with a direct overlapping of O 2$p$ states.

\begin{acknowledgments}
We would like to thank A. Zheludev for a fruitful discussion on the inelastic neutron scattering data. We acknowledge the Helmholtz Virtual Institute (Project No. VH-VI-521).
\end{acknowledgments}



\begin{thebibliography}{41}%
\makeatletter
\providecommand \@ifxundefined [1]{%
 \@ifx{#1\undefined}
}%
\providecommand \@ifnum [1]{%
 \ifnum #1\expandafter \@firstoftwo
 \else \expandafter \@secondoftwo
 \fi
}%
\providecommand \@ifx [1]{%
 \ifx #1\expandafter \@firstoftwo
 \else \expandafter \@secondoftwo
 \fi
}%
\providecommand \natexlab [1]{#1}%
\providecommand \enquote  [1]{``#1''}%
\providecommand \bibnamefont  [1]{#1}%
\providecommand \bibfnamefont [1]{#1}%
\providecommand \citenamefont [1]{#1}%
\providecommand \href@noop [0]{\@secondoftwo}%
\providecommand \href [0]{\begingroup \@sanitize@url \@href}%
\providecommand \@href[1]{\@@startlink{#1}\@@href}%
\providecommand \@@href[1]{\endgroup#1\@@endlink}%
\providecommand \@sanitize@url [0]{\catcode `\\12\catcode `\$12\catcode
  `\&12\catcode `\#12\catcode `\^12\catcode `\_12\catcode `\%12\relax}%
\providecommand \@@startlink[1]{}%
\providecommand \@@endlink[0]{}%
\providecommand \url  [0]{\begingroup\@sanitize@url \@url }%
\providecommand \@url [1]{\endgroup\@href {#1}{\urlprefix }}%
\providecommand \urlprefix  [0]{URL }%
\providecommand \Eprint [0]{\href }%
\providecommand \doibase [0]{http://dx.doi.org/}%
\providecommand \selectlanguage [0]{\@gobble}%
\providecommand \bibinfo  [0]{\@secondoftwo}%
\providecommand \bibfield  [0]{\@secondoftwo}%
\providecommand \translation [1]{[#1]}%
\providecommand \BibitemOpen [0]{}%
\providecommand \bibitemStop [0]{}%
\providecommand \bibitemNoStop [0]{.\EOS\space}%
\providecommand \EOS [0]{\spacefactor3000\relax}%
\providecommand \BibitemShut  [1]{\csname bibitem#1\endcsname}%
\let\auto@bib@innerbib\@empty
\bibitem [{\citenamefont {Affleck}(1989)}]{AffleckJPCM.1.3047}%
  \BibitemOpen
  \bibfield  {author} {\bibinfo {author} {\bibfnamefont {I.}~\bibnamefont
  {Affleck}},\ }\href {\doibase 10.1088/0953-8984/1/19/001} {\bibfield
  {journal} {\bibinfo  {journal} {J. Phys.: Condens. Matter.}\ }\textbf
  {\bibinfo {volume} {1}},\ \bibinfo {pages} {3047} (\bibinfo {year}
  {1989})}\BibitemShut {NoStop}%
\bibitem [{\citenamefont {Smirnov}\ and\ \citenamefont
  {Glazkov}(2007)}]{SmirnovJETP.105.861}%
  \BibitemOpen
  \bibfield  {author} {\bibinfo {author} {\bibfnamefont {A.}~\bibnamefont
  {Smirnov}}\ and\ \bibinfo {author} {\bibfnamefont {V.}~\bibnamefont
  {Glazkov}},\ }\href {\doibase 10.1134/S1063776107100214} {\bibfield
  {journal} {\bibinfo  {journal} {J. Exp. Theore. Phys.}\ }\textbf {\bibinfo
  {volume} {105}},\ \bibinfo {pages} {861} (\bibinfo {year}
  {2007})}\BibitemShut {NoStop}%
\bibitem [{\citenamefont {Haldane}(1983{\natexlab{a}})}]{HaldanePRL.50.1153}%
  \BibitemOpen
  \bibfield  {author} {\bibinfo {author} {\bibfnamefont {F.~D.~M.}\
  \bibnamefont {Haldane}},\ }\href {\doibase 10.1103/PhysRevLett.50.1153}
  {\bibfield  {journal} {\bibinfo  {journal} {Phys. Rev. Lett.}\ }\textbf
  {\bibinfo {volume} {50}},\ \bibinfo {pages} {1153} (\bibinfo {year}
  {1983}{\natexlab{a}})}\BibitemShut {NoStop}%
\bibitem [{\citenamefont {Haldane}(1983{\natexlab{b}})}]{HaldanePLA.93.464}%
  \BibitemOpen
  \bibfield  {author} {\bibinfo {author} {\bibfnamefont {F.~D.~M.}\
  \bibnamefont {Haldane}},\ }\href {\doibase 10.1016/0375-9601(83)90631-X}
  {\bibfield  {journal} {\bibinfo  {journal} {Phys. Lett. A}\ }\textbf
  {\bibinfo {volume} {93}},\ \bibinfo {pages} {464} (\bibinfo {year}
  {1983}{\natexlab{b}})}\BibitemShut {NoStop}%
\bibitem [{\citenamefont {Affleck}\ \emph {et~al.}(1987)\citenamefont
  {Affleck}, \citenamefont {Kennedy}, \citenamefont {Lieb},\ and\ \citenamefont
  {Tasaki}}]{Affleck.PRL.59.799}%
  \BibitemOpen
  \bibfield  {author} {\bibinfo {author} {\bibfnamefont {I.}~\bibnamefont
  {Affleck}}, \bibinfo {author} {\bibfnamefont {T.}~\bibnamefont {Kennedy}},
  \bibinfo {author} {\bibfnamefont {E.~H.}\ \bibnamefont {Lieb}}, \ and\
  \bibinfo {author} {\bibfnamefont {H.}~\bibnamefont {Tasaki}},\ }\href
  {\doibase 10.1103/PhysRevLett.59.799} {\bibfield  {journal} {\bibinfo
  {journal} {Phys. Rev. Lett.}\ }\textbf {\bibinfo {volume} {59}},\ \bibinfo
  {pages} {799} (\bibinfo {year} {1987})}\BibitemShut {NoStop}%
\bibitem [{\citenamefont {Takahashi}(1989)}]{TakahashiPRL.62.2313}%
  \BibitemOpen
  \bibfield  {author} {\bibinfo {author} {\bibfnamefont {M.}~\bibnamefont
  {Takahashi}},\ }\href {\doibase 10.1103/PhysRevLett.62.2313} {\bibfield
  {journal} {\bibinfo  {journal} {Phys. Rev. Lett.}\ }\textbf {\bibinfo
  {volume} {62}},\ \bibinfo {pages} {2313} (\bibinfo {year}
  {1989})}\BibitemShut {NoStop}%
\bibitem [{\citenamefont {Golinelli}\ \emph
  {et~al.}(1993{\natexlab{a}})\citenamefont {Golinelli}, \citenamefont
  {Jolicoeur},\ and\ \citenamefont {Lacaze}}]{Golinelli.JPCM.5.1399}%
  \BibitemOpen
  \bibfield  {author} {\bibinfo {author} {\bibfnamefont {O.}~\bibnamefont
  {Golinelli}}, \bibinfo {author} {\bibfnamefont {T.}~\bibnamefont
  {Jolicoeur}}, \ and\ \bibinfo {author} {\bibfnamefont {R.}~\bibnamefont
  {Lacaze}},\ }\href {http://stacks.iop.org/0953-8984/5/i=9/a=024} {\bibfield
  {journal} {\bibinfo  {journal} {Journal of Physics: Condensed Matter}\
  }\textbf {\bibinfo {volume} {5}},\ \bibinfo {pages} {1399} (\bibinfo {year}
  {1993}{\natexlab{a}})}\BibitemShut {NoStop}%
\bibitem [{\citenamefont {White}\ and\ \citenamefont
  {Affleck}(2008)}]{WhitePRB.77.134437}%
  \BibitemOpen
  \bibfield  {author} {\bibinfo {author} {\bibfnamefont {S.~R.}\ \bibnamefont
  {White}}\ and\ \bibinfo {author} {\bibfnamefont {I.}~\bibnamefont
  {Affleck}},\ }\href {\doibase 10.1103/PhysRevB.77.134437} {\bibfield
  {journal} {\bibinfo  {journal} {Phys. Rev. B}\ }\textbf {\bibinfo {volume}
  {77}},\ \bibinfo {pages} {134437} (\bibinfo {year} {2008})}\BibitemShut
  {NoStop}%
\bibitem [{\citenamefont {Ma}\ \emph {et~al.}(1992)\citenamefont {Ma},
  \citenamefont {Broholm}, \citenamefont {Reich}, \citenamefont {Sternlieb},\
  and\ \citenamefont {Erwin}}]{Ma.PRL.69.3571}%
  \BibitemOpen
  \bibfield  {author} {\bibinfo {author} {\bibfnamefont {S.}~\bibnamefont
  {Ma}}, \bibinfo {author} {\bibfnamefont {C.}~\bibnamefont {Broholm}},
  \bibinfo {author} {\bibfnamefont {D.~H.}\ \bibnamefont {Reich}}, \bibinfo
  {author} {\bibfnamefont {B.~J.}\ \bibnamefont {Sternlieb}}, \ and\ \bibinfo
  {author} {\bibfnamefont {R.~W.}\ \bibnamefont {Erwin}},\ }\href {\doibase
  10.1103/PhysRevLett.69.3571} {\bibfield  {journal} {\bibinfo  {journal}
  {Phys. Rev. Lett.}\ }\textbf {\bibinfo {volume} {69}},\ \bibinfo {pages}
  {3571} (\bibinfo {year} {1992})}\BibitemShut {NoStop}%
\bibitem [{\citenamefont {Xu}\ \emph {et~al.}(2000)\citenamefont {Xu},
  \citenamefont {Aeppli}, \citenamefont {Bisher}, \citenamefont {Broholm},
  \citenamefont {DiTusa}, \citenamefont {Frost}, \citenamefont {Ito},
  \citenamefont {Oka}, \citenamefont {Paul}, \citenamefont {Takagi},\ and\
  \citenamefont {Treacy}}]{Xu.Science.289.419}%
  \BibitemOpen
  \bibfield  {author} {\bibinfo {author} {\bibfnamefont {G.}~\bibnamefont
  {Xu}}, \bibinfo {author} {\bibfnamefont {G.}~\bibnamefont {Aeppli}}, \bibinfo
  {author} {\bibfnamefont {M.~E.}\ \bibnamefont {Bisher}}, \bibinfo {author}
  {\bibfnamefont {C.}~\bibnamefont {Broholm}}, \bibinfo {author} {\bibfnamefont
  {J.~F.}\ \bibnamefont {DiTusa}}, \bibinfo {author} {\bibfnamefont {C.~D.}\
  \bibnamefont {Frost}}, \bibinfo {author} {\bibfnamefont {T.}~\bibnamefont
  {Ito}}, \bibinfo {author} {\bibfnamefont {K.}~\bibnamefont {Oka}}, \bibinfo
  {author} {\bibfnamefont {R.~L.}\ \bibnamefont {Paul}}, \bibinfo {author}
  {\bibfnamefont {H.}~\bibnamefont {Takagi}}, \ and\ \bibinfo {author}
  {\bibfnamefont {M.~M.~J.}\ \bibnamefont {Treacy}},\ }\href {\doibase
  10.1126/science.289.5478.419} {\bibfield  {journal} {\bibinfo  {journal}
  {Science}\ }\textbf {\bibinfo {volume} {289}},\ \bibinfo {pages} {419}
  (\bibinfo {year} {2000})}\BibitemShut {NoStop}%
\bibitem [{\citenamefont {Zaliznyak}\ \emph {et~al.}(2001)\citenamefont
  {Zaliznyak}, \citenamefont {Lee},\ and\ \citenamefont
  {Petrov}}]{ZaliznyakPRL.87.017202}%
  \BibitemOpen
  \bibfield  {author} {\bibinfo {author} {\bibfnamefont {I.~A.}\ \bibnamefont
  {Zaliznyak}}, \bibinfo {author} {\bibfnamefont {S.-H.}\ \bibnamefont {Lee}},
  \ and\ \bibinfo {author} {\bibfnamefont {S.~V.}\ \bibnamefont {Petrov}},\
  }\href {\doibase 10.1103/PhysRevLett.87.017202} {\bibfield  {journal}
  {\bibinfo  {journal} {Phys. Rev. Lett.}\ }\textbf {\bibinfo {volume} {87}},\
  \bibinfo {pages} {017202} (\bibinfo {year} {2001})}\BibitemShut {NoStop}%
\bibitem [{\citenamefont {Sakai}\ and\ \citenamefont
  {Takahashi}(1990)}]{SakaiPRB.42.4537}%
  \BibitemOpen
  \bibfield  {author} {\bibinfo {author} {\bibfnamefont {T.}~\bibnamefont
  {Sakai}}\ and\ \bibinfo {author} {\bibfnamefont {M.}~\bibnamefont
  {Takahashi}},\ }\href {\doibase 10.1103/PhysRevB.42.4537} {\bibfield
  {journal} {\bibinfo  {journal} {Phys. Rev. B}\ }\textbf {\bibinfo {volume}
  {42}},\ \bibinfo {pages} {4537} (\bibinfo {year} {1990})}\BibitemShut
  {NoStop}%
\bibitem [{\citenamefont {Morra}\ \emph {et~al.}(1988)\citenamefont {Morra},
  \citenamefont {Buyers}, \citenamefont {Armstrong},\ and\ \citenamefont
  {Hirakawa}}]{MorraPRB.38.543}%
  \BibitemOpen
  \bibfield  {author} {\bibinfo {author} {\bibfnamefont {R.~M.}\ \bibnamefont
  {Morra}}, \bibinfo {author} {\bibfnamefont {W.~J.~L.}\ \bibnamefont
  {Buyers}}, \bibinfo {author} {\bibfnamefont {R.~L.}\ \bibnamefont
  {Armstrong}}, \ and\ \bibinfo {author} {\bibfnamefont {K.}~\bibnamefont
  {Hirakawa}},\ }\href {\doibase 10.1103/PhysRevB.38.543} {\bibfield  {journal}
  {\bibinfo  {journal} {Phys. Rev. B}\ }\textbf {\bibinfo {volume} {38}},\
  \bibinfo {pages} {543} (\bibinfo {year} {1988})}\BibitemShut {NoStop}%
\bibitem [{\citenamefont {Zheludev}\ \emph {et~al.}(2000)\citenamefont
  {Zheludev}, \citenamefont {Masuda}, \citenamefont {Tsukada}, \citenamefont
  {Uchiyama}, \citenamefont {Uchinokura}, \citenamefont {B\"oni},\ and\
  \citenamefont {Lee}}]{ZheludevPRB.62.8921}%
  \BibitemOpen
  \bibfield  {author} {\bibinfo {author} {\bibfnamefont {A.}~\bibnamefont
  {Zheludev}}, \bibinfo {author} {\bibfnamefont {T.}~\bibnamefont {Masuda}},
  \bibinfo {author} {\bibfnamefont {I.}~\bibnamefont {Tsukada}}, \bibinfo
  {author} {\bibfnamefont {Y.}~\bibnamefont {Uchiyama}}, \bibinfo {author}
  {\bibfnamefont {K.}~\bibnamefont {Uchinokura}}, \bibinfo {author}
  {\bibfnamefont {P.}~\bibnamefont {B\"oni}}, \ and\ \bibinfo {author}
  {\bibfnamefont {S.-H.}\ \bibnamefont {Lee}},\ }\href {\doibase
  10.1103/PhysRevB.62.8921} {\bibfield  {journal} {\bibinfo  {journal} {Phys.
  Rev. B}\ }\textbf {\bibinfo {volume} {62}},\ \bibinfo {pages} {8921}
  (\bibinfo {year} {2000})}\BibitemShut {NoStop}%
\bibitem [{\citenamefont {Zheludev}\ \emph {et~al.}(2001)\citenamefont
  {Zheludev}, \citenamefont {Masuda}, \citenamefont {Uchinokura},\ and\
  \citenamefont {Nagler}}]{ZheludevPRB.64.134415}%
  \BibitemOpen
  \bibfield  {author} {\bibinfo {author} {\bibfnamefont {A.}~\bibnamefont
  {Zheludev}}, \bibinfo {author} {\bibfnamefont {T.}~\bibnamefont {Masuda}},
  \bibinfo {author} {\bibfnamefont {K.}~\bibnamefont {Uchinokura}}, \ and\
  \bibinfo {author} {\bibfnamefont {S.~E.}\ \bibnamefont {Nagler}},\ }\href
  {\doibase 10.1103/PhysRevB.64.134415} {\bibfield  {journal} {\bibinfo
  {journal} {Phys. Rev. B}\ }\textbf {\bibinfo {volume} {64}},\ \bibinfo
  {pages} {134415} (\bibinfo {year} {2001})}\BibitemShut {NoStop}%
\bibitem [{\citenamefont {Pahari}\ \emph {et~al.}(2006)\citenamefont {Pahari},
  \citenamefont {Ghoshray}, \citenamefont {Sarkar}, \citenamefont
  {Bandyopadhyay},\ and\ \citenamefont {Ghoshray}}]{PahariPRB.73.012407}%
  \BibitemOpen
  \bibfield  {author} {\bibinfo {author} {\bibfnamefont {B.}~\bibnamefont
  {Pahari}}, \bibinfo {author} {\bibfnamefont {K.}~\bibnamefont {Ghoshray}},
  \bibinfo {author} {\bibfnamefont {R.}~\bibnamefont {Sarkar}}, \bibinfo
  {author} {\bibfnamefont {B.}~\bibnamefont {Bandyopadhyay}}, \ and\ \bibinfo
  {author} {\bibfnamefont {A.}~\bibnamefont {Ghoshray}},\ }\href {\doibase
  10.1103/PhysRevB.73.012407} {\bibfield  {journal} {\bibinfo  {journal} {Phys.
  Rev. B}\ }\textbf {\bibinfo {volume} {73}},\ \bibinfo {pages} {012407}
  (\bibinfo {year} {2006})}\BibitemShut {NoStop}%
\bibitem [{\citenamefont {Bera}\ \emph {et~al.}(2013)\citenamefont {Bera},
  \citenamefont {Lake}, \citenamefont {Islam}, \citenamefont {Klemke},
  \citenamefont {Faulhaber},\ and\ \citenamefont {Law}}]{BeraPRB.87.224423}%
  \BibitemOpen
  \bibfield  {author} {\bibinfo {author} {\bibfnamefont {A.~K.}\ \bibnamefont
  {Bera}}, \bibinfo {author} {\bibfnamefont {B.}~\bibnamefont {Lake}}, \bibinfo
  {author} {\bibfnamefont {A.~T. M.~N.}\ \bibnamefont {Islam}}, \bibinfo
  {author} {\bibfnamefont {B.}~\bibnamefont {Klemke}}, \bibinfo {author}
  {\bibfnamefont {E.}~\bibnamefont {Faulhaber}}, \ and\ \bibinfo {author}
  {\bibfnamefont {J.~M.}\ \bibnamefont {Law}},\ }\href {\doibase
  10.1103/PhysRevB.87.224423} {\bibfield  {journal} {\bibinfo  {journal} {Phys.
  Rev. B}\ }\textbf {\bibinfo {volume} {87}},\ \bibinfo {pages} {224423}
  (\bibinfo {year} {2013})}\BibitemShut {NoStop}%
\bibitem [{\citenamefont {Konik}\ and\ \citenamefont
  {Fendley}(2002)}]{KonikPRB.66.144416}%
  \BibitemOpen
  \bibfield  {author} {\bibinfo {author} {\bibfnamefont {R.~M.}\ \bibnamefont
  {Konik}}\ and\ \bibinfo {author} {\bibfnamefont {P.}~\bibnamefont
  {Fendley}},\ }\href {\doibase 10.1103/PhysRevB.66.144416} {\bibfield
  {journal} {\bibinfo  {journal} {Phys. Rev. B}\ }\textbf {\bibinfo {volume}
  {66}},\ \bibinfo {pages} {144416} (\bibinfo {year} {2002})}\BibitemShut
  {NoStop}%
\bibitem [{\citenamefont {Affleck}(1992)}]{AffleckPRB.46.9002}%
  \BibitemOpen
  \bibfield  {author} {\bibinfo {author} {\bibfnamefont {I.}~\bibnamefont
  {Affleck}},\ }\href {\doibase 10.1103/PhysRevB.46.9002} {\bibfield  {journal}
  {\bibinfo  {journal} {Phys. Rev. B}\ }\textbf {\bibinfo {volume} {46}},\
  \bibinfo {pages} {9002} (\bibinfo {year} {1992})}\BibitemShut {NoStop}%
\bibitem [{\citenamefont {Farutin}\ and\ \citenamefont
  {Marchenko}(2007)}]{FarutinJEPT.104.751}%
  \BibitemOpen
  \bibfield  {author} {\bibinfo {author} {\bibfnamefont {A.}~\bibnamefont
  {Farutin}}\ and\ \bibinfo {author} {\bibfnamefont {V.}~\bibnamefont
  {Marchenko}},\ }\href {\doibase 10.1134/S1063776107050093} {\bibfield
  {journal} {\bibinfo  {journal} {J. Exp. Theo. Phys.}\ }\textbf {\bibinfo
  {volume} {104}},\ \bibinfo {pages} {751} (\bibinfo {year}
  {2007})}\BibitemShut {NoStop}%
\bibitem [{\citenamefont {Golinelli}\ \emph
  {et~al.}(1993{\natexlab{b}})\citenamefont {Golinelli}, \citenamefont
  {Jolicoeur},\ and\ \citenamefont {Lacaze}}]{GolinelliJPCM.5.7847}%
  \BibitemOpen
  \bibfield  {author} {\bibinfo {author} {\bibfnamefont {O.}~\bibnamefont
  {Golinelli}}, \bibinfo {author} {\bibfnamefont {T.}~\bibnamefont
  {Jolicoeur}}, \ and\ \bibinfo {author} {\bibfnamefont {R.}~\bibnamefont
  {Lacaze}},\ }\href {http://stacks.iop.org/0953-8984/5/i=42/a=007} {\bibfield
  {journal} {\bibinfo  {journal} {J. Phys.: Condens. Matter}\ }\textbf
  {\bibinfo {volume} {5}},\ \bibinfo {pages} {7847} (\bibinfo {year}
  {1993}{\natexlab{b}})}\BibitemShut {NoStop}%
\bibitem [{\citenamefont {Regnault}\ \emph {et~al.}(1994)\citenamefont
  {Regnault}, \citenamefont {Zaliznyak}, \citenamefont {Renard},\ and\
  \citenamefont {Vettier}}]{Regnault.PRB.50.9174}%
  \BibitemOpen
  \bibfield  {author} {\bibinfo {author} {\bibfnamefont {L.~P.}\ \bibnamefont
  {Regnault}}, \bibinfo {author} {\bibfnamefont {I.}~\bibnamefont {Zaliznyak}},
  \bibinfo {author} {\bibfnamefont {J.~P.}\ \bibnamefont {Renard}}, \ and\
  \bibinfo {author} {\bibfnamefont {C.}~\bibnamefont {Vettier}},\ }\href
  {\doibase 10.1103/PhysRevB.50.9174} {\bibfield  {journal} {\bibinfo
  {journal} {Phys. Rev. B}\ }\textbf {\bibinfo {volume} {50}},\ \bibinfo
  {pages} {9174} (\bibinfo {year} {1994})}\BibitemShut {NoStop}%
\bibitem [{\citenamefont {Zheludev}\ \emph {et~al.}(2003)\citenamefont
  {Zheludev}, \citenamefont {Honda}, \citenamefont {Broholm}, \citenamefont
  {Katsumata}, \citenamefont {Shapiro}, \citenamefont {Kolezhuk}, \citenamefont
  {Park},\ and\ \citenamefont {Qiu}}]{Zheludev.PRB.68.134438}%
  \BibitemOpen
  \bibfield  {author} {\bibinfo {author} {\bibfnamefont {A.}~\bibnamefont
  {Zheludev}}, \bibinfo {author} {\bibfnamefont {Z.}~\bibnamefont {Honda}},
  \bibinfo {author} {\bibfnamefont {C.~L.}\ \bibnamefont {Broholm}}, \bibinfo
  {author} {\bibfnamefont {K.}~\bibnamefont {Katsumata}}, \bibinfo {author}
  {\bibfnamefont {S.~M.}\ \bibnamefont {Shapiro}}, \bibinfo {author}
  {\bibfnamefont {A.}~\bibnamefont {Kolezhuk}}, \bibinfo {author}
  {\bibfnamefont {S.}~\bibnamefont {Park}}, \ and\ \bibinfo {author}
  {\bibfnamefont {Y.}~\bibnamefont {Qiu}},\ }\href {\doibase
  10.1103/PhysRevB.68.134438} {\bibfield  {journal} {\bibinfo  {journal} {Phys.
  Rev. B}\ }\textbf {\bibinfo {volume} {68}},\ \bibinfo {pages} {134438}
  (\bibinfo {year} {2003})}\BibitemShut {NoStop}%
\bibitem [{\citenamefont {Smirnov}\ \emph {et~al.}(2008)\citenamefont
  {Smirnov}, \citenamefont {Glazkov}, \citenamefont {Kashiwagi}, \citenamefont
  {Kimura}, \citenamefont {Hagiwara}, \citenamefont {Kindo}, \citenamefont
  {Shapiro},\ and\ \citenamefont {Demianets}}]{SmirnovPRB.77.100401}%
  \BibitemOpen
  \bibfield  {author} {\bibinfo {author} {\bibfnamefont {A.~I.}\ \bibnamefont
  {Smirnov}}, \bibinfo {author} {\bibfnamefont {V.~N.}\ \bibnamefont
  {Glazkov}}, \bibinfo {author} {\bibfnamefont {T.}~\bibnamefont {Kashiwagi}},
  \bibinfo {author} {\bibfnamefont {S.}~\bibnamefont {Kimura}}, \bibinfo
  {author} {\bibfnamefont {M.}~\bibnamefont {Hagiwara}}, \bibinfo {author}
  {\bibfnamefont {K.}~\bibnamefont {Kindo}}, \bibinfo {author} {\bibfnamefont
  {A.~Y.}\ \bibnamefont {Shapiro}}, \ and\ \bibinfo {author} {\bibfnamefont
  {L.~N.}\ \bibnamefont {Demianets}},\ }\href {\doibase
  10.1103/PhysRevB.77.100401} {\bibfield  {journal} {\bibinfo  {journal} {Phys.
  Rev. B}\ }\textbf {\bibinfo {volume} {77}},\ \bibinfo {pages} {100401}
  (\bibinfo {year} {2008})}\BibitemShut {NoStop}%
\bibitem [{\citenamefont {Islam}\ and\ \citenamefont
  {et~al.}(2015)}]{Islam.unpublished}%
  \BibitemOpen
  \bibfield  {author} {\bibinfo {author} {\bibfnamefont {A.~T. M.~N.}\
  \bibnamefont {Islam}}\ and\ \bibinfo {author} {\bibnamefont {et~al.}},\
  }\href@noop {} {} (\bibinfo {year} {2015}),\ \Eprint
  {http://arxiv.org/abs/unpublished} {unpublished} \BibitemShut {NoStop}%
\bibitem [{Ful()}]{Fullprof}%
  \BibitemOpen
  \href@noop {} {\enquote {\bibinfo {title} {Fullprof suite},}\ }\bibinfo
  {howpublished} {http://www.ill.eu/sites/fullprof/}\BibitemShut {NoStop}%
\bibitem [{\citenamefont {Koepernik}\ and\ \citenamefont
  {Eschrig}(1999)}]{dft:fplo}%
  \BibitemOpen
  \bibfield  {author} {\bibinfo {author} {\bibfnamefont {K.}~\bibnamefont
  {Koepernik}}\ and\ \bibinfo {author} {\bibfnamefont {H.}~\bibnamefont
  {Eschrig}},\ }\href {\doibase 10.1103/PhysRevB.59.1743} {\bibfield  {journal}
  {\bibinfo  {journal} {Phys. Rev. B}\ }\textbf {\bibinfo {volume} {59}},\
  \bibinfo {pages} {1743} (\bibinfo {year} {1999})}\BibitemShut {NoStop}%
\bibitem [{\citenamefont {Perdew}\ \emph {et~al.}(1996)\citenamefont {Perdew},
  \citenamefont {Burke},\ and\ \citenamefont {Ernzerhof}}]{dft:pbe96}%
  \BibitemOpen
  \bibfield  {author} {\bibinfo {author} {\bibfnamefont {J.~P.}\ \bibnamefont
  {Perdew}}, \bibinfo {author} {\bibfnamefont {K.}~\bibnamefont {Burke}}, \
  and\ \bibinfo {author} {\bibfnamefont {M.}~\bibnamefont {Ernzerhof}},\ }\href
  {\doibase 10.1103/PhysRevLett.77.3865} {\bibfield  {journal} {\bibinfo
  {journal} {Phys. Rev. Lett.}\ }\textbf {\bibinfo {volume} {77}},\ \bibinfo
  {pages} {3865} (\bibinfo {year} {1996})}\BibitemShut {NoStop}%
\bibitem [{\citenamefont {Eschrig}\ and\ \citenamefont
  {Koepernik}(2009)}]{dft:fplo_wf}%
  \BibitemOpen
  \bibfield  {author} {\bibinfo {author} {\bibfnamefont {H.}~\bibnamefont
  {Eschrig}}\ and\ \bibinfo {author} {\bibfnamefont {K.}~\bibnamefont
  {Koepernik}},\ }\href {\doibase 10.1103/PhysRevB.80.104503} {\bibfield
  {journal} {\bibinfo  {journal} {Phys. Rev. B}\ }\textbf {\bibinfo {volume}
  {80}},\ \bibinfo {pages} {104503} (\bibinfo {year} {2009})}\BibitemShut
  {NoStop}%
\bibitem [{\citenamefont {Bera}\ and\ \citenamefont
  {Yusuf}(2012)}]{BeraPRB.86.024408}%
  \BibitemOpen
  \bibfield  {author} {\bibinfo {author} {\bibfnamefont {A.~K.}\ \bibnamefont
  {Bera}}\ and\ \bibinfo {author} {\bibfnamefont {S.~M.}\ \bibnamefont
  {Yusuf}},\ }\href {\doibase 10.1103/PhysRevB.86.024408} {\bibfield  {journal}
  {\bibinfo  {journal} {Phys. Rev. B}\ }\textbf {\bibinfo {volume} {86}},\
  \bibinfo {pages} {024408} (\bibinfo {year} {2012})}\BibitemShut {NoStop}%
\bibitem [{\citenamefont {Wichmann}\ and\ \citenamefont
  {M{\"u}ller-Buschbaum}(1986)}]{dft:str}%
  \BibitemOpen
  \bibfield  {author} {\bibinfo {author} {\bibfnamefont {R.}~\bibnamefont
  {Wichmann}}\ and\ \bibinfo {author} {\bibfnamefont {H.}~\bibnamefont
  {M{\"u}ller-Buschbaum}},\ }\href@noop {} {\bibfield  {journal} {\bibinfo
  {journal} {Rev. Chim. Mineral.}\ }\textbf {\bibinfo {volume} {23}},\ \bibinfo
  {pages} {1} (\bibinfo {year} {1986})}\BibitemShut {NoStop}%
\bibitem [{\citenamefont {Regnault}\ \emph {et~al.}(1993)\citenamefont
  {Regnault}, \citenamefont {Zaliznyak},\ and\ \citenamefont
  {Meshkov}}]{RegnaultJPCM.5.L677}%
  \BibitemOpen
  \bibfield  {author} {\bibinfo {author} {\bibfnamefont {L.~P.}\ \bibnamefont
  {Regnault}}, \bibinfo {author} {\bibfnamefont {I.~A.}\ \bibnamefont
  {Zaliznyak}}, \ and\ \bibinfo {author} {\bibfnamefont {S.~V.}\ \bibnamefont
  {Meshkov}},\ }\href {http://stacks.iop.org/0953-8984/5/i=50/a=004} {\bibfield
   {journal} {\bibinfo  {journal} {J. Phys.: Condens. Matter}\ }\textbf
  {\bibinfo {volume} {5}},\ \bibinfo {pages} {L677} (\bibinfo {year}
  {1993})}\BibitemShut {NoStop}%
\bibitem [{\citenamefont {Golinelli}\ \emph {et~al.}(1992)\citenamefont
  {Golinelli}, \citenamefont {Jolicoeur},\ and\ \citenamefont
  {Lacaze}}]{GolinelliPRB.46.10854}%
  \BibitemOpen
  \bibfield  {author} {\bibinfo {author} {\bibfnamefont {O.}~\bibnamefont
  {Golinelli}}, \bibinfo {author} {\bibfnamefont {T.}~\bibnamefont
  {Jolicoeur}}, \ and\ \bibinfo {author} {\bibfnamefont {R.}~\bibnamefont
  {Lacaze}},\ }\href {\doibase 10.1103/PhysRevB.46.10854} {\bibfield  {journal}
  {\bibinfo  {journal} {Phys. Rev. B}\ }\textbf {\bibinfo {volume} {46}},\
  \bibinfo {pages} {10854} (\bibinfo {year} {1992})}\BibitemShut {NoStop}%
\bibitem [{\citenamefont {Wang}\ \emph {et~al.}(2013)\citenamefont {Wang},
  \citenamefont {Schmidt}, \citenamefont {Bera}, \citenamefont {Islam},
  \citenamefont {Lake}, \citenamefont {Loidl},\ and\ \citenamefont
  {Deisenhofer}}]{WangPRB.87.104405}%
  \BibitemOpen
  \bibfield  {author} {\bibinfo {author} {\bibfnamefont {Z.}~\bibnamefont
  {Wang}}, \bibinfo {author} {\bibfnamefont {M.}~\bibnamefont {Schmidt}},
  \bibinfo {author} {\bibfnamefont {A.~K.}\ \bibnamefont {Bera}}, \bibinfo
  {author} {\bibfnamefont {A.~T. M.~N.}\ \bibnamefont {Islam}}, \bibinfo
  {author} {\bibfnamefont {B.}~\bibnamefont {Lake}}, \bibinfo {author}
  {\bibfnamefont {A.}~\bibnamefont {Loidl}}, \ and\ \bibinfo {author}
  {\bibfnamefont {J.}~\bibnamefont {Deisenhofer}},\ }\href {\doibase
  10.1103/PhysRevB.87.104405} {\bibfield  {journal} {\bibinfo  {journal} {Phys.
  Rev. B}\ }\textbf {\bibinfo {volume} {87}},\ \bibinfo {pages} {104405}
  (\bibinfo {year} {2013})}\BibitemShut {NoStop}%
\bibitem [{\citenamefont {Goodenough}(1955)}]{Goodenough.PR.100.564}%
  \BibitemOpen
  \bibfield  {author} {\bibinfo {author} {\bibfnamefont {J.~B.}\ \bibnamefont
  {Goodenough}},\ }\href {\doibase 10.1103/PhysRev.100.564} {\bibfield
  {journal} {\bibinfo  {journal} {Phys. Rev.}\ }\textbf {\bibinfo {volume}
  {100}},\ \bibinfo {pages} {564} (\bibinfo {year} {1955})}\BibitemShut
  {NoStop}%
\bibitem [{\citenamefont {Kanamori}(1959)}]{Kanamori.JPCS.10.87}%
  \BibitemOpen
  \bibfield  {author} {\bibinfo {author} {\bibfnamefont {J.}~\bibnamefont
  {Kanamori}},\ }\href {\doibase
  http://dx.doi.org/10.1016/0022-3697(59)90061-7} {\bibfield  {journal}
  {\bibinfo  {journal} {J. Phys. Chem. Solids}\ }\textbf {\bibinfo {volume}
  {10}},\ \bibinfo {pages} {87 } (\bibinfo {year} {1959})}\BibitemShut
  {NoStop}%
\bibitem [{\citenamefont {M{\"{o}}ller}\ \emph {et~al.}(2009)\citenamefont
  {M{\"{o}}ller}, \citenamefont {Schmitt}, \citenamefont {Schnelle},
  \citenamefont {F{\"{o}}rster},\ and\ \citenamefont
  {Rosner}}]{dft:moeller2009}%
  \BibitemOpen
  \bibfield  {author} {\bibinfo {author} {\bibfnamefont {A.}~\bibnamefont
  {M{\"{o}}ller}}, \bibinfo {author} {\bibfnamefont {M.}~\bibnamefont
  {Schmitt}}, \bibinfo {author} {\bibfnamefont {W.}~\bibnamefont {Schnelle}},
  \bibinfo {author} {\bibfnamefont {T.}~\bibnamefont {F{\"{o}}rster}}, \ and\
  \bibinfo {author} {\bibfnamefont {H.}~\bibnamefont {Rosner}},\ }\href
  {\doibase 10.1103/PhysRevB.80.125106} {\bibfield  {journal} {\bibinfo
  {journal} {Phys. Rev. B}\ }\textbf {\bibinfo {volume} {80}},\ \bibinfo
  {pages} {125106} (\bibinfo {year} {2009})}\BibitemShut {NoStop}%
\bibitem [{\citenamefont {Tsirlin}\ \emph {et~al.}(2010)\citenamefont
  {Tsirlin}, \citenamefont {Janson},\ and\ \citenamefont
  {Rosner}}]{dft:tsirlin2010}%
  \BibitemOpen
  \bibfield  {author} {\bibinfo {author} {\bibfnamefont {A.~A.}\ \bibnamefont
  {Tsirlin}}, \bibinfo {author} {\bibfnamefont {O.}~\bibnamefont {Janson}}, \
  and\ \bibinfo {author} {\bibfnamefont {H.}~\bibnamefont {Rosner}},\ }\href
  {\doibase 10.1103/PhysRevB.82.144416} {\bibfield  {journal} {\bibinfo
  {journal} {Phys. Rev. B}\ }\textbf {\bibinfo {volume} {82}},\ \bibinfo
  {pages} {144416} (\bibinfo {year} {2010})}\BibitemShut {NoStop}%
\bibitem [{\citenamefont {Janson}\ \emph {et~al.}(2011)\citenamefont {Janson},
  \citenamefont {Tsirlin}, \citenamefont {Sichelschmidt}, \citenamefont
  {Skourski}, \citenamefont {Weickert},\ and\ \citenamefont
  {Rosner}}]{dft:janson2011}%
  \BibitemOpen
  \bibfield  {author} {\bibinfo {author} {\bibfnamefont {O.}~\bibnamefont
  {Janson}}, \bibinfo {author} {\bibfnamefont {A.~A.}\ \bibnamefont {Tsirlin}},
  \bibinfo {author} {\bibfnamefont {J.}~\bibnamefont {Sichelschmidt}}, \bibinfo
  {author} {\bibfnamefont {Y.}~\bibnamefont {Skourski}}, \bibinfo {author}
  {\bibfnamefont {F.}~\bibnamefont {Weickert}}, \ and\ \bibinfo {author}
  {\bibfnamefont {H.}~\bibnamefont {Rosner}},\ }\href {\doibase
  10.1103/PhysRevB.83.094435} {\bibfield  {journal} {\bibinfo  {journal} {Phys.
  Rev. B}\ }\textbf {\bibinfo {volume} {83}},\ \bibinfo {pages} {094435}
  (\bibinfo {year} {2011})}\BibitemShut {NoStop}%
\bibitem [{\citenamefont {Arango}\ \emph {et~al.}(2011)\citenamefont {Arango},
  \citenamefont {Vavilova}, \citenamefont {Abdel-Hafiez}, \citenamefont
  {Janson}, \citenamefont {Tsirlin}, \citenamefont {Rosner}, \citenamefont
  {Drechsler}, \citenamefont {Weil}, \citenamefont {N{\'e}nert}, \citenamefont
  {Klingeler}, \citenamefont {Volkova}, \citenamefont {Vasiliev}, \citenamefont
  {Kataev},\ and\ \citenamefont {B{\"u}chner}}]{dft:arango2011}%
  \BibitemOpen
  \bibfield  {author} {\bibinfo {author} {\bibfnamefont {Y.~C.}\ \bibnamefont
  {Arango}}, \bibinfo {author} {\bibfnamefont {E.}~\bibnamefont {Vavilova}},
  \bibinfo {author} {\bibfnamefont {M.}~\bibnamefont {Abdel-Hafiez}}, \bibinfo
  {author} {\bibfnamefont {O.}~\bibnamefont {Janson}}, \bibinfo {author}
  {\bibfnamefont {A.~A.}\ \bibnamefont {Tsirlin}}, \bibinfo {author}
  {\bibfnamefont {H.}~\bibnamefont {Rosner}}, \bibinfo {author} {\bibfnamefont
  {S.-L.}\ \bibnamefont {Drechsler}}, \bibinfo {author} {\bibfnamefont
  {M.}~\bibnamefont {Weil}}, \bibinfo {author} {\bibfnamefont {G.}~\bibnamefont
  {N{\'e}nert}}, \bibinfo {author} {\bibfnamefont {R.}~\bibnamefont
  {Klingeler}}, \bibinfo {author} {\bibfnamefont {O.}~\bibnamefont {Volkova}},
  \bibinfo {author} {\bibfnamefont {A.}~\bibnamefont {Vasiliev}}, \bibinfo
  {author} {\bibfnamefont {V.}~\bibnamefont {Kataev}}, \ and\ \bibinfo {author}
  {\bibfnamefont {B.}~\bibnamefont {B{\"u}chner}},\ }\href {\doibase
  10.1103/PhysRevB.84.134430} {\bibfield  {journal} {\bibinfo  {journal} {Phys.
  Rev. B}\ }\textbf {\bibinfo {volume} {84}},\ \bibinfo {pages} {134430}
  (\bibinfo {year} {2011})}\BibitemShut {NoStop}%
\bibitem [{\citenamefont {Nishimoto}\ \emph {et~al.}(2012)\citenamefont
  {Nishimoto}, \citenamefont {Drechsler}, \citenamefont {Kuzian}, \citenamefont
  {Richter}, \citenamefont {M\'alek}, \citenamefont {Schmitt}, \citenamefont
  {van~den Brink},\ and\ \citenamefont {Rosner}}]{Nishimoto.EPL.98.37007}%
  \BibitemOpen
  \bibfield  {author} {\bibinfo {author} {\bibfnamefont {S.}~\bibnamefont
  {Nishimoto}}, \bibinfo {author} {\bibfnamefont {S.-L.}\ \bibnamefont
  {Drechsler}}, \bibinfo {author} {\bibfnamefont {R.}~\bibnamefont {Kuzian}},
  \bibinfo {author} {\bibfnamefont {J.}~\bibnamefont {Richter}}, \bibinfo
  {author} {\bibfnamefont {J.}~\bibnamefont {M\'alek}}, \bibinfo {author}
  {\bibfnamefont {M.}~\bibnamefont {Schmitt}}, \bibinfo {author} {\bibfnamefont
  {J.}~\bibnamefont {van~den Brink}}, \ and\ \bibinfo {author} {\bibfnamefont
  {H.}~\bibnamefont {Rosner}},\ }\href {\doibase 10.1209/0295-5075/98/37007}
  {\bibfield  {journal} {\bibinfo  {journal} {EuroPhys. Lett.}\ }\textbf
  {\bibinfo {volume} {98}},\ \bibinfo {pages} {37007} (\bibinfo {year}
  {2012})}\BibitemShut {NoStop}%
\end{thebibliography}
%


\end{document}